\documentclass[twocolumn, resetfootnote]{aastex7}

\def\OI{[\mbox{O\,{\sc i}}]$\lambda 6300$}
\def\OIII{[\mbox{O\,{\sc iii}}]$\lambda 5007$}

\def\SIIab{[\mbox{S\,{\sc ii}}]$\lambda\lambda 6717,6731$}
\def\SII{[\mbox{S\,{\sc ii}}]$\lambda \lambda 6717,6731$}
\def\NII{[\mbox{N\,{\sc ii}}]$\lambda 6584$}
\def\NIIb{[\mbox{N\,{\sc ii}}]$\lambda 6584$}

\def\NIIab{[\mbox{N\,{\sc  ii}}]$\lambda \lambda 6547,6584$}

\def\OI{[\mbox{O{\sc i}}]$\lambda 6300$}

\def\Ha{{H$\alpha$}}
\def\Hb{{H$\beta$}}

\def\NIIHa{[\mbox{N\,{\sc ii}}]$\lambda 6583$/H$\alpha$}
\def\SIIHa{[\mbox{S\,{\sc ii}}]$\lambda\lambda 6717,6731$/H$\alpha$}

\def\OIHa{[\mbox{O\,{\sc i}}]$\lambda 6300$/H$\alpha$}
\def\OIIIHb{[\mbox{O\,{\sc iii}}]$\lambda 5007$/H$\beta$}

\def\LOIIIs4{$L[\mbox{O\,{\sc iii}}]$/$\sigma^4$}

\def\kms{${\rm km}~{\rm s}^{-1}$}
\newcommand{\ergcms}	{\ifmmode {\rm erg\,cm}^{-2}\,{\rm s}^{-1} \else erg\,cm$^{-2}$\,s$^{-1}$\fi}

\newcommand{\gandalf}{{\texttt {gandalf}}}
 
\newcommand{\ppxf}{{\texttt {pPXF}}}

\definecolor{myblue}{RGB}{0, 100, 220}
\definecolor{myred}{RGB}{225, 0, 100}

\usepackage{graphicx}
\usepackage{sistyle}
\SIthousandsep{,}
\usepackage{multirow}

\received{April 27, 2026}
\revised{June 9, 2026}
\accepted{June 10, 2026}
\journalinfo{}

\begin{document}

\title{Decoupled Kinematics and Excitation in the Compton-thick AGN NGC 6552: Spatially Resolved KOOLS-IFU Observations}

\author[orcid=0000-0002-5037-951X,sname='Oh']{Kyuseok Oh}
\affiliation{Korea Astronomy and Space Science Institute, Daedeok-daero 776, Yuseong-gu, Daejeon 34055, Republic of Korea}
\email[show]{oh@kasi.re.kr}  
\correspondingauthor{Kyuseok Oh}

\author[orcid=0000-0001-7821-6715,sname='Ueda']{Yoshihiro Ueda}
\affiliation{Department of Astronomy, Kyoto University, Kitashirakawa-Oiwake-cho, Sakyo-ku, Kyoto, 606-8502, Japan}
\email{ueda@kusastro.kyoto-u.ac.jp}  

\author[orcid=0009-0006-4377-4219,sname='Fujiwara']{Kanta Fujiwara}
\affiliation{Department of Astronomy, Kyoto University, Kitashirakawa-Oiwake-cho, Sakyo-ku, Kyoto, 606-8502, Japan}
\email{fujiwara@kusastro.kyoto-u.ac.jp}  

\author[orcid=0000-0002-6480-3799,sname='Isogai']{Keisuke Isogai}
\affiliation{Okayama Observatory, Kyoto University, 3037-5 Honjo, Kamogatacho, Asakuchi, Okayama 719-0232, Japan}
\affiliation{Department of Multi-Disciplinary Sciences, Graduate School of Arts and Sciences, The University of Tokyo, 3-8-1 Komaba, Meguro, Tokyo 153-8902, Japan}
\email{isogai@kusastro.kyoto-u.ac.jp}  

\author[orcid=0000-0002-9754-3081,sname='Yamada']{Satoshi Yamada}
\affiliation{Frontier Research Institute for Interdisciplinary Sciences, Tohoku University, Sendai, Miyagi 980-8578, Japan}
\affiliation{Department of Astronomy, University of Geneva, Ch.d’Ecogia 16, 1290, Versoix, Switzerland.}
\affiliation{Astronomical Institute, Tohoku University, 6-3 Aramakiazaaoba, Aoba-ku, Sendai, Miyagi 980-8578, Japan}
\email{satoshi.yamada@astr.tohoku.ac.jp}

\author[orcid=0009-0005-5861-1263,sname='Shimoda']{Keito Shimoda}
\affiliation{Department of Astronomy, Kyoto University, Kitashirakawa-Oiwake-cho, Sakyo-ku, Kyoto, 606-8502, Japan}
\email{shimoda@kusastro.kyoto-u.ac.jp}  

\author[orcid=0000-0002-3531-7863, sname='Toba']{Yoshiki Toba}
\affiliation{Department of Physical Sciences, Ritsumeikan University, 1-1-1 Noji-higashi, Kusatsu, Shiga 525-8577, Japan}
\affiliation{Research Center for Space and Cosmic Evolution, Ehime University, 2-5 Bunkyo-cho, Matsuyama, Ehime 790-8577, Japan}
\affiliation{Academia Sinica Institute of Astronomy and Astrophysics, 11F of ASMAB, No.~1, Section 4, Roosevelt Road, Taipei 10617, Taiwan}
\email{toba@fc.ritsumei.ac.jp}  

\author[orcid=0000-0001-6473-5100,sname='Matsubayashi']{Kazuya Matsubayashi}
\affiliation{Institute of Astronomy, The University of Tokyo, 2-21-1 Osawa, Mitaka, Tokyo 181-0015, Japan}
\email{kazuya@ioa.s.u-tokyo.ac.jp}  

\author[orcid=0000-0002-5701-0811,sname='Ogawa']{Shoji Ogawa}
\affiliation{Faculty of Science and Technology, Tokyo University of Science, 2641 Yamazaki, Noda, 2788510, Chiba, Japan}
\email{ogawa.shoji@rs.tus.ac.jp}  

\author[orcid=0009-0000-9577-8701,sname='Nakatani']{Yuya Nakatani}
\affiliation{Department of Astronomy, Kyoto University, Kitashirakawa-Oiwake-cho, Sakyo-ku, Kyoto, 606-8502, Japan}
\email{nakatani@kusastro.kyoto-u.ac.jp}

\begin{abstract}
Hard X-ray selected Compton-thick AGNs provide a relatively obscuration-resistant census of accretion, 
but optical line diagnostics can be strongly shaped by extinction and geometry. 
Spatially resolved integral-field spectroscopy can mitigate these effects and provides 
direct constraints on outflow kinematics and ionization state on kiloparsec scales.
We present KOOLS-IFU optical integral-field spectroscopy of NGC\,6552 obtained on the 3.8\,m Seimei Telescope. 
Using spatially resolved emission-line ratios and non-parametric \OIII\ kinematics over the central $\sim2$\,kpc, 
we test whether ionized-gas kinematics are locally coupled to excitation. 
The \OIII\ width $W_{80}$ is broadly elevated across the inner region ($\sim$530--830~km~s$^{-1}$) and 
declines monotonically with projected galactocentric distance, 
consistent with a centrally concentrated outflow that decelerates at larger radii. 
Despite this clear kinematic structure, neither $W_{80}$ nor the velocity asymmetry parameter $\Delta v$ shows 
a statistically significant correlation with \OIIIHb. 
Order-of-magnitude outflow energetics yield $\dot{E}_{K}/L_{\rm bol} \approx 0.01\%$--$0.28\%$ 
(for assumed $n_{e} = 50$--$1000$~cm$^{-3}$), 
consistent with $[\mbox{O\,{\sc iii}}]$-based estimates tracing only the ionized phase of a multi-phase outflow. 
We conclude that in NGC~6552 both the total line broadening traced by $W_{80}$ and $\Delta v$ are 
consistent with being governed primarily by spatial dynamical structure and 
line-of-sight superposition of multiple kinematic components, 
with no statistically significant coupling to excitation-driven processes detected at our sensitivity level. 
A positive $W_{80}$--\OIIIHb\ coupling does emerge in the small subset of bins for which the two-component fit is most strongly favored statistically, 
which deeper observations will be needed to confirm.
\end{abstract}

\keywords{\uat{Active galactic nuclei}{16} --- \uat{AGN host galaxies}{2017} --- \uat{Seyfert galaxies}{1447} --- \uat{Emission line galaxies}{459}}



\section{Introduction}\label{sec:intro}

Active galactic nuclei (AGNs) selected at hard X-ray energies provide 
a relatively obscuration-resistant view of accretion activity, 
including heavily obscured and Compton-thick systems \citep{Burlon11, Ueda14, Koss16, Ricci17}. 
Because these sources can host powerful winds while their optical emission may be 
strongly affected by extinction and anisotropic illumination, 
multi-wavelength approaches are essential for characterizing AGN-driven outflows and 
their potential impact on host galaxies \citep{Fabian12, Harrison18}.

Optical integral-field spectroscopy (IFS) offers spatially resolved constraints on 
ionized-gas kinematics and excitation on kiloparsec (kpc) scales. 
Resolved studies frequently reveal broad emission-line wings, 
velocity asymmetries, and radially varying behaviors in 
non-parametric velocity measures such as $W_{80}$, 
often interpreted as signatures of outflow--ISM interaction \citep{Liu13, Harrison14, McElroy15, Kakkad22, Gatto24, Oh25}. 
In parallel, classical emission-line diagnostics (e.g., \OIIIHb\ and \NIIHa) are widely used 
to infer ionization mechanisms and to separate AGN photoionization, 
star formation, and shock-like excitation or systematic shifts in Baldwin, Phillips \& Terlevich \citep[BPT,][]{Baldwin81} space 
\citep{Kewley01, Kewley06, Kauffmann03, Schawinski07, Allen08}.

A key open question is whether kinematic disturbance and excitation state are spatially coupled on resolved scales. 
If outflow-driven shocks contribute substantially to the line emission, 
regions with large velocity widths might exhibit line-ratio signatures consistent with shock-like excitation or 
systematic shifts in BPT space \citep{Allen08, Rich11, Rich15}. 
Furthermore, shock-driven line ratios partly overlap with the photoionization-driven predictions in the BPT space, 
limiting the ability of BPT diagnostics alone to disentangle the two ionization mechanisms \citep[e.g.,][]{Allen08, DAgostino19a}.
However, projection effects, density structure, and line-of-sight superposition of multiple ionization components 
can complicate or even erase a one-to-one correspondence \citep[e.g.,][]{Fischer13}. 
This issue is especially relevant for Compton-thick AGNs, 
where extinction and anisotropic obscuration can bias optical diagnostics 
toward the less-obscured regions of the narrow-line emitting gas \citep{Netzer15, Hickox18}.

NGC~6552 is a nearby ($z=0.0260$) hard X-ray selected AGN \citep{Oh18, Lien25}. 
The source is heavily obscured, with $\log(N_{\rm H}/\mathrm{cm}^{-2}) = 24.05$ \citep{Ricci17}, 
placing it at the Compton-thick threshold. 
An earlier measurement based on 0.7--10~keV data yielded a lower value of $\log(N_{\rm H}/\mathrm{cm}^{-2}) = 23.78$ \citep{Fukazawa94}, 
likely reflecting the difficulty of constraining $N_{\rm H}$ without hard X-ray coverage above 10~keV. 
Recent JWST/MIRI MRS mid-infrared integral-field spectroscopy provided the first clear evidence for a nuclear outflow in this system, 
detected as blue-shifted velocity components in high-excitation 
and coronal emission lines spanning ionization potentials of 27--187~eV \citep{AlvarezMarquez23}. 
However, this mid-infrared outflow was spatially unresolved and confined to $< 0.2$~kpc from the nucleus. 
Whether the outflow propagates to larger scales and produces detectable signatures in the optical \OIII\ emission remained untested, 
motivating a complementary optical IFU observation that can probe kpc-scale kinematics in the lower-density, 
less-obscured ionized gas traced by \OIII\ and test whether the kinematic disturbance is spatially coupled to excitation diagnostics.

For context, we summarize basic AGN properties of NGC\,6552 from the literature. 
Based on the stellar velocity dispersion ($\sigma_\ast = 202.04 \pm 9.54$~km~s$^{-1}$) 
reported by \citet{Koss22_velocity_dispersion}, 
they derive $\log(M_{\rm BH}/M_\odot)=8.51^{+0.09}_{-0.09}$. 
They also report $\log L_{\rm bol}=44.51$ (erg~s$^{-1}$) and $\log(L_{\rm bol}/L_{\rm Edd})=-2.18$ \citep{Koss22_catalog_and_data}. 
These values are provided for context and are not required for our kinematics--excitation analysis.

In this work, we present KOOLS-IFU observations of NGC\,6552 obtained on the 3.8\,m Seimei Telescope. 
We map spatially resolved \OIII\ kinematics using non-parametric velocity measures and 
construct emission-line ratio maps within the central $\sim2$\,kpc. 
We then test the relationship between kinematic broadening and ionization state 
using correlation, partial-correlation, and radial-trend analyses. 
We also provide order-of-magnitude estimates of the ionized outflow energetics 
to place NGC~6552 in the context of AGN feedback scaling relations. 
The remainder of this paper is organized as follows. 
Section~\ref{sec:obs} describes the observations and analysis. 
Section~\ref{sec:obs_emlines} describes the spectral fitting and kinematic measurements. 
Section~\ref{sec:results} presents the results. Section~\ref{sec:discussion} discusses implications and caveats. 
Section~\ref{sec:summary} summarizes our conclusions. 
Throughout this paper, we adopt a cosmology with $h=0.70$, $\Omega_M=0.30$, and $\Omega_\Lambda=0.70$.

\begin{figure}
\centering
\includegraphics[width=0.90\linewidth]{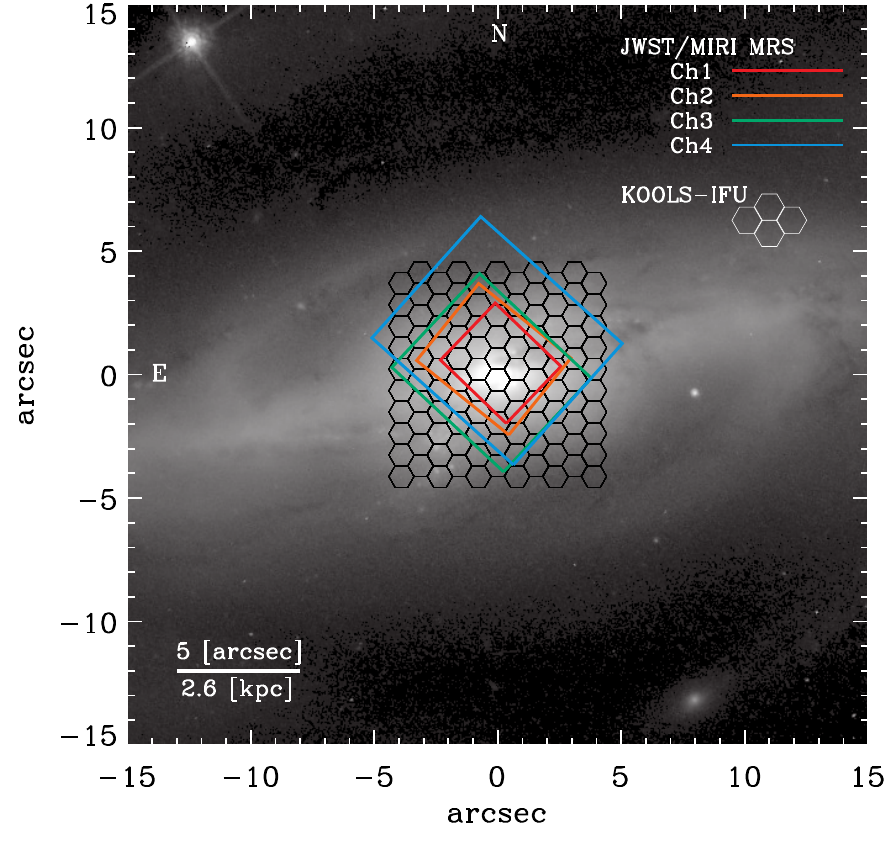}
\caption{HST/ACS F814W image of the Compton-thick AGN host galaxy NGC\,6552 \citep{Kim21}. 
The black hexagons represent the KOOLS-IFU fiber layout and cover the central $\sim8.4^{\prime\prime} \times 8.0^{\prime\prime}$ region 
(corresponding to $\sim4.3\times4.1$~kpc at $z=0.0260$). 
The colored boxes indicate the footprints of the four channels of the JWST/MIRI MRS observations reported by 
\citet{AlvarezMarquez23}, covering progressively larger fields of view from Ch1 ($\sim$4.9--7.7~$\mu$m) to Ch4 
($\sim$17.7--27.9~$\mu$m).
This overlay is shown to provide the nuclear mid-infrared context relative to the larger optical IFU coverage used in this work.
North is up and east is to the left. The scale bar indicates $5^{\prime\prime} \approx 2.6$~kpc.
\label{fig:NGC6552}}
\end{figure}

\begin{deluxetable*}{lcccccccc}
\tablecaption{Target information and observation log.\label{tab:obslog_allinone}}
\tablehead{
\multirow{2}{*}{Name} &
\multirow{2}{*}{R.A.} &
\multirow{2}{*}{Decl.} &
\multirow{2}{*}{$z$} &
\multirow{2}{*}{AGN type} &
\multirow{2}{*}{Date} &
\multicolumn{2}{c}{Exp.\ time} &
\multirow{2}{*}{Seeing} \\
 & & & & & & VPH495 & VPH683 & 
}
\startdata
NGC\,6552 & 18:00:07.27 & +66:36:54.33 & 0.026 & Sy 2 & 2025.03.30 &
600$\times$3~s$^{*}$ & 600$\times$3~s$^{*}$ & $1.30^{\prime\prime}$  \\
\enddata
\tablecomments{Coordinates are given in the J2000 equatorial system.
$^{*}$Total integration time is $600\times3$~s (1800~s) for each grism.
Spectrophotometric calibration was performed 
using the standard star HR\,5501.}
\end{deluxetable*}

\section{Observations and Data Reduction}\label{sec:obs}

\subsection{KOOLS-IFU Observations}\label{sec:obs_kools}

The observational data were obtained with the Kyoto Okayama Optical Low-dispersion Spectrograph fiber IFU instrument 
\citep[KOOLS-IFU,][]{Matsubayashi19, Matsubayashi25} on the 3.8\,m Seimei Telescope \citep{Kurita20} under the programme (25A-K-0019, PI: Y. Ueda). 
KOOLS-IFU consists of 110 fibers with a total field of view of $8.4^{\prime\prime} \times 8.0^{\prime\prime}$, 
corresponding to $\sim4.3\times4.1$~kpc at $z=0.0260$ (Figure~\ref{fig:NGC6552}). 
We used the VPH 495 grism (4300--5900~\AA, $R \equiv \lambda/\Delta\lambda \approx 1500$, 
velocity FWHM $\sim 200$~km~s$^{-1}$) and the VPH 683 grism (5800--8000~\AA, $R \approx 2000$, velocity FWHM $\sim 150$~km~s$^{-1}$).

Observing conditions were assessed from the spectrophotometric standard-star exposures. 
For each grism, we constructed a fiber-sampled spatial profile of the standard star by computing 
a continuum flux for each fiber, measured as the median flux within two continuum windows near \OIII\ 
(4890--4920\,\AA\ and 5025--5055\,\AA\ in the reduced spectra). 
These windows were chosen to lie close to \OIII\ while avoiding strong nebular emission features. 
We then estimated the point-spread function (PSF) size from the flux-weighted spatial second moments 
of the fiber flux distribution and converted the resulting dispersions to an effective circularized FWHM. 
The inferred seeing is consistent between the two grisms, with a representative value of 
$\mathrm{FWHM}\simeq1.3^{\prime\prime}$ during our observations.
The observation log (dates, exposure times, and observing conditions) is summarized in Table~\ref{tab:obslog_allinone}.

\subsection{Data Reduction}\label{sec:obs_reduction}

We reduced the data using the standard KOOLS-IFU reduction pipeline\footnote{\url{https://www.o.kwasan.kyoto-u.ac.jp/inst/p-kools/reduction-201806/index.html}}, 
which employs the Image Reduction and Analysis Facility (IRAF, \citealt{Tody86, Tody93}). 
For spectrum extraction, flat-fielding, and wavelength calibration, we used the Hydra package \citep{Barden94, Barden95}. 
We calibrated the wavelength using arc-lamp frames (Ne and Hg), and performed absolute flux calibration using 
spectrophotometric standard star observations. 
To account for the background sky level, we obtained sky frames at a nearby blank sky field with the same exposure time 
(600~s per frame, 3 frames per grism) as the science observations.

Considering the typical seeing at the site ($\sim1.2^{\prime\prime}$--$1.4^{\prime\prime}$) and the field of view of each fiber 
(a regular hexagon with radius $0.42^{\prime\prime}$), we combined data from three adjacent fibers and present the results 
throughout this work.

The KOOLS-IFU footprint is aligned to the HST/ACS F814W image (Figure~\ref{fig:NGC6552}). 
The nuclear position adopted throughout this work is defined as the centroid of the background-subtracted JWST/MIRI 
F560W image reported by \citet{AlvarezMarquez23}. 
This position is used as the reference center for projected galactocentric distances 
(marked by a cross in the kinematic and line-ratio maps presented in Section~\ref{sec:results}).

\begin{figure*}
    \centering
    \includegraphics[width=0.95\linewidth]{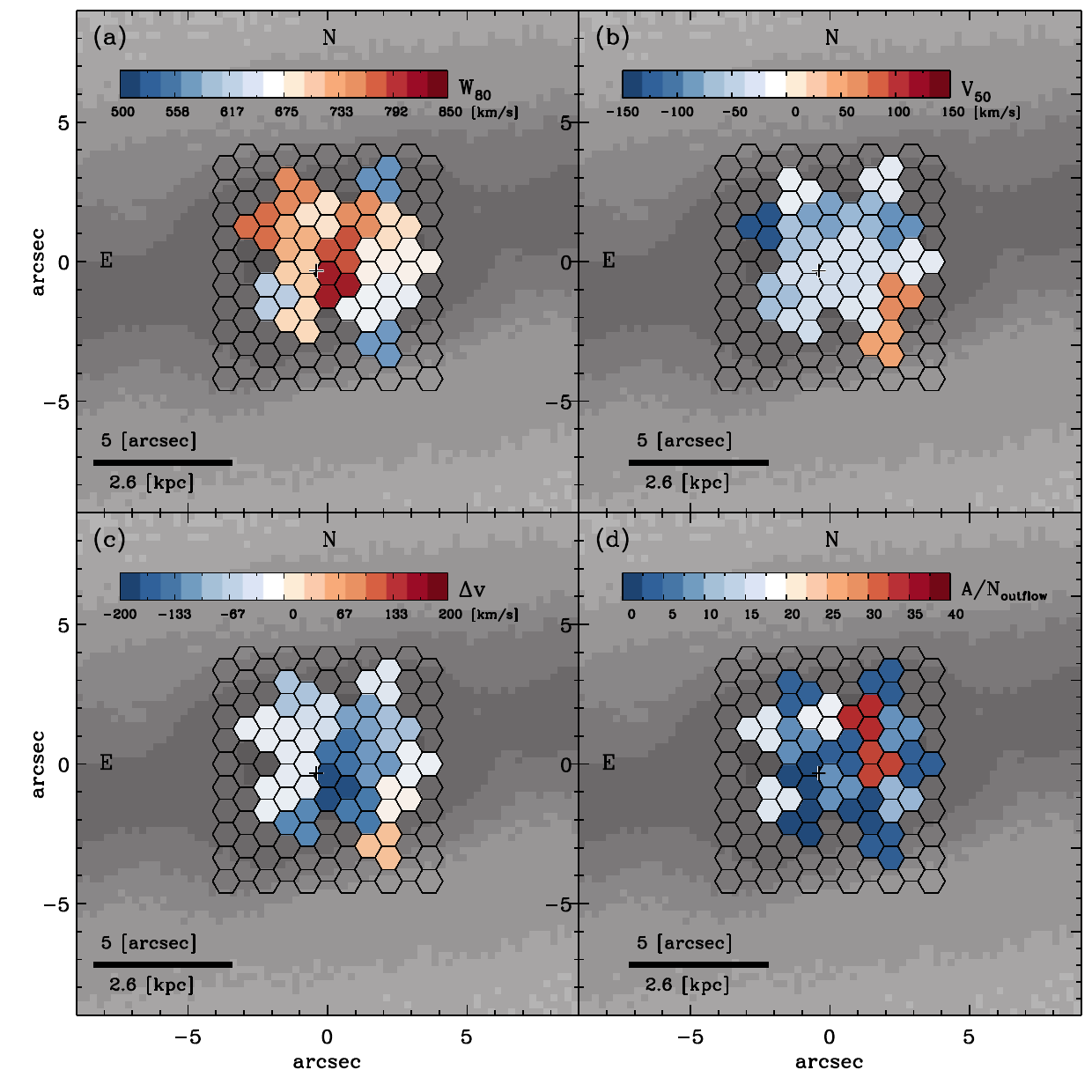}
    \caption{
    Spatially resolved \OIII\ kinematics of the AGN host galaxy NGC~6552 derived from KOOLS-IFU observations.
    The background greyscale image is the Pan-STARRS1 $r$-band image \citep{Chambers16}.
    All panels show bins where the shifted \OIII\ component is detected with 
    an amplitude-to-noise ratio A/N$>2$, 
    emphasizing the spatial distribution of asymmetric line wings.
    The nucleus position (cross) is defined from the JWST/MIRI F560W centroid \citep{AlvarezMarquez23}.
    (a) Non-parametric velocity width $W_{80}=v_{90}-v_{10}$ (km~s$^{-1}$), tracing the overall line broadening including
    high-velocity wings.
    Elevated $W_{80}$ values extend to $\sim$1.5~kpc from the nucleus, indicating kpc-scale ionized-gas kinematic
    disturbance.
    (b) Median velocity $v_{50}$ (km~s$^{-1}$) relative to the systemic velocity.
    The color scale is centered at 0~km~s$^{-1}$ to highlight blue- and redshifted regions symmetrically.
    The moderate velocity amplitudes suggest that bulk rotation alone does not account for the enhanced line widths.
    (c) Velocity asymmetry parameter $\Delta v=(v_{05}+v_{95})/2 - v_{50}$ (km~s$^{-1}$), which quantifies the relative
    dominance of blue or red wings.
    The color scale is centered at 0~km~s$^{-1}$ to highlight blueshifted- and redshifted-wing-dominated regions symmetrically.
    Negative (positive) values indicate stronger blue (red) line wings, consistent with asymmetric outflow signatures.
    (d) Amplitude-to-noise ratio map of the shifted \OIII\ outflow component (see Section~\ref{sec:obs_outflow_sn} for the definition). 
    The shifted component provides a parametric description of the asymmetric \OIII\ line wings.
    \label{fig:OIII_kinematics}}
\end{figure*}

\section{Spectral Fitting and Kinematic Measurements}\label{sec:obs_emlines}

\subsection{Spectral Fitting}\label{sec:spec_fitting}
We performed spectral fitting following the procedures described by \citet{Sarzi06}, 
which have been extensively applied to spectroscopic data ranging from large SDSS galaxy samples \citep{Oh11}\footnote{\url{https://data.kasi.re.kr/vo/OSSY/}} 
to optical follow-up spectroscopy of the Swift-BAT ultra-hard X-ray all-sky survey \citep{Oh17, Oh22}\footnote{\url{https://bass-survey.com}}.
We first deredshifted the extracted spectra and corrected for Galactic foreground extinction using 
${\rm E(B-V)}=0.0451$ \citep{Schlegel98} with the attenuation curve of \citet{Calzetti00}. 

We measured the stellar kinematics (line-of-sight velocity and velocity dispersion) 
using the penalized pixel fitting method (\ppxf; \citealt{Cappellari04}), 
with emission-line spectral regions masked using windows of 1200~\kms\ width. 
We adopted stellar population models from \citet{Bruzual03} and the MILES empirical stellar libraries 
\citep{SanchezBlazquez06}. 
We also masked prominent sky lines (5577\,\AA, 6300\,\AA, 6363\,\AA) and 
the NaD $\lambda\lambda$5890,5896 absorption lines with the same mask width.

We then used the \gandalf\ code \citep{Sarzi06} to simultaneously fit the stellar continuum 
and the nebular emission lines, with the stellar line-of-sight velocity dispersion fixed 
to the value measured in the \ppxf\ step and used to broaden the stellar templates. 
This two-step approach separates the optimization of stellar kinematics, 
for which \ppxf\ is specifically designed, 
from the multi-line emission fitting that \gandalf\ implements 
\citep[see also][]{Sarzi06, Oh11}.

All Balmer emission lines were modeled with narrow components only, 
as no broad Balmer features (FWHM $\gtrsim 1000$~\kms) are detected in NGC\,6552, 
consistent with its obscured (Compton-thick) AGN classification.

For the \OIII\ emission, we adopted a two-component Gaussian decomposition 
using a core component plus an additional shifted component 
representing a kinematically shifted, outflow-related contribution. 
In each spatial bin, the core component has one free parameter per line (amplitude), 
with its velocity and width tied to the common forbidden-line kinematics. 
The forbidden-line velocity and dispersion are themselves free parameters of the fit 
in each spatial bin, with the initial velocity guess taken from the \ppxf-derived 
stellar line-of-sight velocity in the same bin. This procedure naturally accommodates 
bulk galaxy rotation through the spatially varying stellar kinematics.
The shifted component has three free parameters: amplitude, velocity dispersion, and velocity offset relative to the core. 
The velocity offset of the shifted component is left as a free parameter.
Unlike the core component, the shifted component is fitted as an independent free parameter and is not kinematically tied to other forbidden lines. 
The decomposition of the \OIII\ profile into core and shifted components is therefore determined solely by the \OIII\ line profile, 
independent of the kinematics of \NII\ or \SII.

The two-component (core + shifted) decomposition is applied only to the \OIII\ emission. 
All other emission lines (e.g., \Ha, \Hb, \NIIab, and \SIIab) are modeled with a single core component only, 
with their velocity and dispersion tied to the common forbidden- or Balmer-line kinematics described above. 
We did not attempt to fit a shifted (outflow) component to lines other than \OIII, 
because the spectral resolution of the present KOOLS-IFU data is not sufficient to reliably separate a narrow core 
and a kinematically shifted component in those lines.

We adopted the two-component ($\mathrm{core}+\mathrm{shift}$) model for \OIII\ only when the shifted component is detected 
with a Gaussian amplitude-to-noise ratio A/N$>2$, where A is the peak amplitude (flux density at the Gaussian peak) 
of the shifted component and N is the local continuum noise level. 
The amplitude-to-noise ratio (A/N) used in this work is distinct from the more familiar signal-to-noise ratio (S/N).  
A/N is defined as the peak flux density (amplitude) of the fitted Gaussian component divided by the local continuum noise, 
whereas S/N typically refers to the integrated line flux divided by the integrated noise. 
The A/N is the standard quality metric for assessing the detection significance of individual Gaussian components 
in multi-component emission-line fits \citep[e.g.,][]{Sarzi06}. 
We use A/N for assessing the detection of the shifted \OIII\ component throughout this work, 
while S/N is reserved for describing the overall line or continuum detection quality.
While A/N$>3$ is the more commonly adopted detection threshold in GANDALF-based analyses \citep[e.g.,][]{Sarzi06, Oh11}, 
A/N$>2$ has also been employed in large spectroscopic surveys \citep{Thomas13}. 
We adopt the lower threshold here because applying A/N$>3$ removes a bin near the nucleus, 
producing an artificial hole in the kinematic maps and degrading their interpretability (Section~\ref{sec:obs_quality}). 
Otherwise, we retained a single core component.

In this paper, we use the term ``shifted (outflow) component'' to refer to the additional \OIII\ Gaussian component 
offset in velocity from the systemic core. 
We interpret this component as outflow-related, and we use it as a parametric description of the asymmetric \OIII\ line wings. 
This component should not be confused with broad permitted-line emission from the broad-line region.

In the simultaneous emission-line fitting, we tied the velocity dispersions within line families to reduce degeneracy. 
Forbidden lines were fit with a common $\sigma$ within each kinematic component. 
Balmer lines were fit with a separate common $\sigma$ within each kinematic component. 
Thus, the core component has one $\sigma$ for forbidden lines and one $\sigma$ for Balmer lines. 
The shifted component, when present, is parameterized in the same way. 
Relative line strengths for selected transitions were adopted based on atomic physics and gas temperature 
(e.g., doublets and Balmer line ratios; see Table~1 in \citealt{Oh11}). 
To determine the shifts and widths of the Gaussian templates, 
we used the standard Levenberg--Marquardt optimization (\texttt{MPFIT} IDL routine; \citealt{Markwardt09}). 
The stellar line-of-sight velocity dispersion derived from the \ppxf\ step was used to broaden the stellar templates in the joint fit.

Emission-line flux uncertainties were estimated by resampling each spectrum with 
100 noise realizations and taking the 1$\sigma$ dispersion of the recovered flux distribution. 
We use the same realizations to propagate uncertainties to the non-parametric kinematic measures described below.

\subsection{Quality cuts and sample definitions}\label{sec:obs_quality}
To ensure robust measurements, we apply quality cuts tailored to each diagnostic, 
because the required signal differs between detecting the shifted \OIII\ component, 
constructing line-ratio maps, and performing correlation tests. 
For the \OIII\ kinematic maps in Figure~\ref{fig:OIII_kinematics}, 
we display spatial bins where the shifted \OIII\ (outflow) Gaussian component is detected with A/N$>2$. 
Using a more conservative threshold (e.g., A/N$>3$) removes only two bins, but one of them lies near the nucleus. 
This produces an artificial ``hole'' in the map and degrades the visual continuity 
and interpretability of the two-dimensional kinematic maps. 
We therefore adopt A/N$>2$ for Figure~\ref{fig:OIII_kinematics} to retain a visually continuous map 
while still requiring a clear detection of the shifted component (see Appendix~\ref{app:specfit_examples} for 
representative spectral profiles across all spatial bins).
To verify that this selection criterion does not bias the apparent spatial distribution of elevated $W_{80}$ values, 
we present in Appendix~\ref{app:w80_selection} a $W_{80}$ map constructed 
without imposing any detection requirement on the shifted component (Figure~\ref{fig:W80_map_all}), 
confirming that the main kinematic features are unchanged.

For emission-line ratio maps, we require A/N$>3$ in both lines used in the ratio (Section~\ref{sec:results_excitation}). 
BPT-based classification additionally requires A/N$>3$ in all four lines (\NII, \Ha, \OIII, and \Hb), 
which naturally yields a more compact classification footprint that is primarily limited by the S/N of H$\beta$. 
The radial $W_{80}(r)$ profile (Section~\ref{sec:results_w80_radial}) uses all bins with measured $W_{80}$ values 
(20 bins), without imposing an additional A/N threshold beyond the requirement for reliable \OIII\ profile measurements. 
Finally, the kinematics--excitation correlation tests (Section~\ref{sec:results_coupling}) are restricted 
to bins satisfying A/N$>3$ in both \OIII\ and \Hb\ ($N_{\rm bins}=15$), 
because the line ratio $\log([\mathrm{O\,III}]/\mathrm{H}\beta)$ requires robust measurements of both lines.

\subsection{Non-parametric velocity measures}\label{sec:obs_nonparam}
We characterize the \OIII\ kinematics using non-parametric velocity measures based on the best-fit emission-line model profile in velocity space. 
For bins fitted with two components, the model profile is the sum of the core and shifted Gaussian components. 
Using the best-fit model profile rather than the raw spectrum reduces the sensitivity of the percentile velocities 
(particularly $v_5$, $v_{10}$, $v_{90}$, and $v_{95}$ in the line wings) to continuum noise and residual contamination, 
and ensures that the measurement is performed on a continuum-subtracted, emission-only profile. 
This approach is the standard practice in non-parametric AGN outflow kinematics analyses \citep[e.g.,][]{Liu13, Harrison14, Kakkad22}.
We note that the percentile velocities probing the high-velocity wings ($v_{5}$, $v_{10}$, $v_{90}$, and $v_{95}$), and hence $W_{80}$ and 
$\Delta v$, are dominated by the shifted component, whose kinematics are fitted independently of the other forbidden lines (Section~\ref{sec:spec_fitting}). 
The kinematic tying of the \OIII\ \emph{core} to the common forbidden-line component therefore has only a limited effect on the non-parametric measures, 
which are set primarily by the untied shifted component in the line wings rather than by the tied core near the systemic velocity.
For each spatial bin, we compute the velocities $v_{p}$ that enclose $p$\% of the total line flux for
$p = 5, 10, 50, 90,$ and $95$.
The percentile velocities are measured from the sampled velocity--flux curve using interpolation.

We define the median velocity as $v_{50}$ and the non-parametric velocity width as
\begin{equation}
W_{80} \equiv v_{90} - v_{10},
\end{equation}
which traces the overall line broadening including the line wings (Figure~\ref{fig:OIII_kinematics}a).
To quantify profile asymmetry, we define
\begin{equation}
\Delta v \equiv \frac{v_{05}+v_{95}}{2} - v_{50},
\end{equation}
where negative (positive) $\Delta v$ indicates a stronger blue (red) wing (Figure~\ref{fig:OIII_kinematics}c). 
The median velocity $v_{50}$ is mapped in Figure~\ref{fig:OIII_kinematics}b.  
Uncertainties on $W_{80}$, $v_{50}$, and $\Delta v$ are estimated via Monte Carlo propagation using the same 100 noise
realizations described above, and we adopt the 1$\sigma$ dispersion as the measurement uncertainty.

Given the KOOLS-IFU instrumental velocity FWHM stated in Section~\ref{sec:obs_kools} ($\sim 150$--$200$~km~s$^{-1}$), 
instrumental broadening is not expected to dominate the observed \OIII\ widths ($W_{80} \sim 530$--$830$~km~s$^{-1}$), 
although it may contribute at a modest level. 
Even without an explicit instrumental correction, 
this would not change the qualitative radial trend or the conclusion that $W_{80}$ is dominated by astrophysical broadening.

An instrumental correction applied in quadrature would change $W_{80}$ by an amount that is small compared to the
observed dynamic range across the field.
We therefore report the observed $W_{80}$ values for consistency across spatial bins.

\subsection{Outflow amplitude-to-noise map}\label{sec:obs_outflow_sn}

To assess the robustness of the shifted \OIII\ component, 
we define an \OIII\ outflow A/N ratio for each spatial bin based on the two-component Gaussian decomposition. 
The outflow A/N is computed as the best-fit amplitude of the shifted (blue- or red-shifted) \OIII\ component divided by the local noise level. 
The noise is estimated from the continuum RMS measured in line-free spectral regions near \OIII. 
The resulting outflow A/N map is shown in Figure~\ref{fig:OIII_kinematics}d.

\subsection{Bayesian Information Criterion Analysis of the Two-Component Fit}
\label{sec:bic}

To complement the A/N threshold adopted for the shifted \OIII\ component (Section~\ref{sec:obs_quality}), 
we performed a Bayesian Information Criterion (BIC; \citealt{Schwarz78}) analysis 
to quantify the statistical preference for the two-component over the single-component \OIII\ fit 
on a bin-by-bin basis, following the approach applied to emission-line decomposition by \citet{Arnaudova24}.
For each spatial bin we computed
\begin{equation}
\Delta\mathrm{BIC} \equiv \mathrm{BIC}_{1} - \mathrm{BIC}_{2}
= \chi^{2}_{1} - \chi^{2}_{2} - \Delta k \ln N_{\mathrm{pix}},
\end{equation}
where $\chi^{2}_{1}$ and $\chi^{2}_{2}$ are the fit residuals of the one- and two-component models 
within a velocity window of $\pm 2400$~km~s$^{-1}$ around the systemic \OIII\ wavelength, 
$\Delta k = 3$ is the difference in the number of free parameters 
between the two models (amplitude, velocity dispersion, and velocity offset of the shifted component), 
and $N_{\mathrm{pix}}$ is the number of spectral pixels in the fitting window. 
Positive $\Delta$BIC values favor the two-component model. 
We adopt the evidence categories of \citet{Kass95}: 
$\Delta\mathrm{BIC} > 0$ (any evidence), 
$\Delta\mathrm{BIC} > 6$ (strong evidence), and 
$\Delta\mathrm{BIC} > 10$ (very strong evidence). 
The spatial distribution of $\Delta$BIC is presented in Appendix~\ref{app:appendix_bic} (Figure~\ref{fig:dBIC_map}), 
and the resulting BIC-based subsample analysis of the kinematics--excitation correlations is given in the same appendix.

\section{Results}\label{sec:results}

The KOOLS-IFU field (Figure~\ref{fig:NGC6552}) covers the central $\sim8.4^{\prime\prime} \times 8.0^{\prime\prime}$ region of NGC\,6552 ($\sim4.3\times4.1$~kpc at $z=0.0260$). 
Bins with sufficient continuum S/N for emission-line analysis are concentrated within the central $\sim2$~kpc, 
enabling a spatially resolved view of ionized gas on kpc scales. 
Projected galactocentric distances are measured relative to the nucleus position defined from the background-subtracted 
JWST/MIRI F560W image \citep{AlvarezMarquez23} (cross in Figures~\ref{fig:OIII_kinematics}).

\subsection{Measurement Coverage and Line-detection Limits}\label{sec:results_coverage}

The spatial coverage of derived quantities is not uniform because each diagnostic requires different emission lines and signal-to-noise thresholds. 
The \OIII\ kinematic maps are derived for bins with reliable \OIII\ measurements (Figure~\ref{fig:OIII_kinematics}), 
and the outflow A/N map (Figure~\ref{fig:OIII_kinematics}d) highlights where the shifted-component signal is robust. 

For excitation diagnostics, line-ratio maps are constructed only for bins that satisfy A/N$>3$ in the required line pair (Section~\ref{sec:results_excitation}). 
In practice, H$\beta$ sets the limiting footprint for $\log([\mathrm{O\,III}]\lambda5007/\mathrm{H}\beta)$, 
and the set of bins that can be classified on the BPT diagram (requiring A/N$>3$ in \NII, \Ha, \OIII, and \Hb) is 
therefore compact around the nucleus. 
These coverage effects are important for interpreting the spatial comparison between kinematics and excitation, 
and for the statistical tests in Section~\ref{sec:results_coupling}.

\begin{figure}
    \centering
    \includegraphics[width=0.95\linewidth]{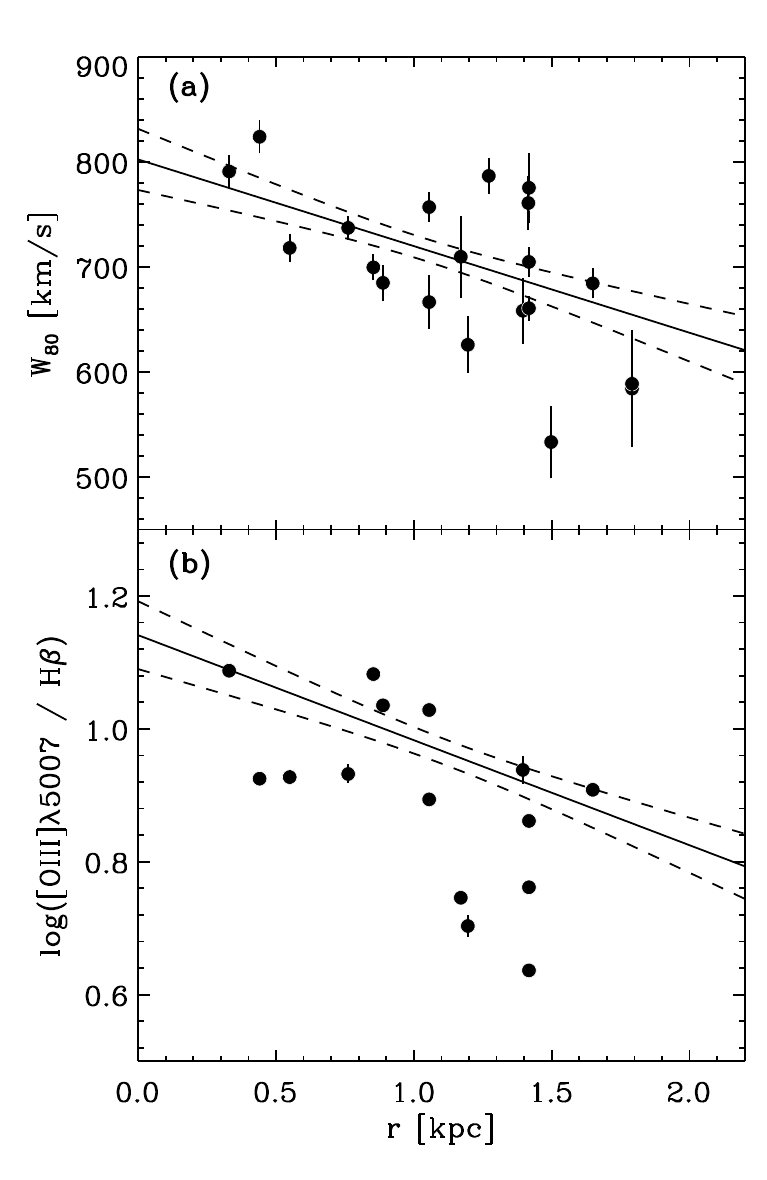}
\caption{Radial profiles of (a) the \OIII\ non-parametric velocity width $W_{80}$ 
and (b) the excitation diagnostic $\log([\mathrm{O\,III}]/\mathrm{H}\beta)$ 
as a function of projected galactocentric distance $r$ (kpc) from the JWST F560W-defined nucleus. 
Each point represents a spatial bin, 
and the vertical error bars show Monte Carlo uncertainties. 
Panel (a) includes all $N=20$ bins with measured $W_{80}$, 
while panel (b) includes the $N=15$ bins satisfying A/N $> 3$ in both \OIII\ and \Hb\ 
(the same sample used for the kinematics--excitation correlation tests of Section~\ref{sec:results_coupling}). 
The solid lines show weighted linear fits, $y = a + br$, 
and dashed curves indicate the $1\sigma$ confidence bands from the parameter covariance matrices. 
For $W_{80}$, an $F$-test against a quadratic model yields $p = 0.80$, 
indicating that the linear model is sufficient. 
The reduced chi-square of the $W_{80}$ linear fit is $\chi^{2}_{\nu} \approx 7.4$, 
indicating substantial intrinsic scatter around the smooth trend. 
Both quantities show statistically significant negative correlations with $r$ 
(Table~\ref{tab:corr_tests}).
    \label{fig:W80_vs_radius}}
\end{figure}

\begin{figure*}
    \centering
    \includegraphics[width=0.95\linewidth]{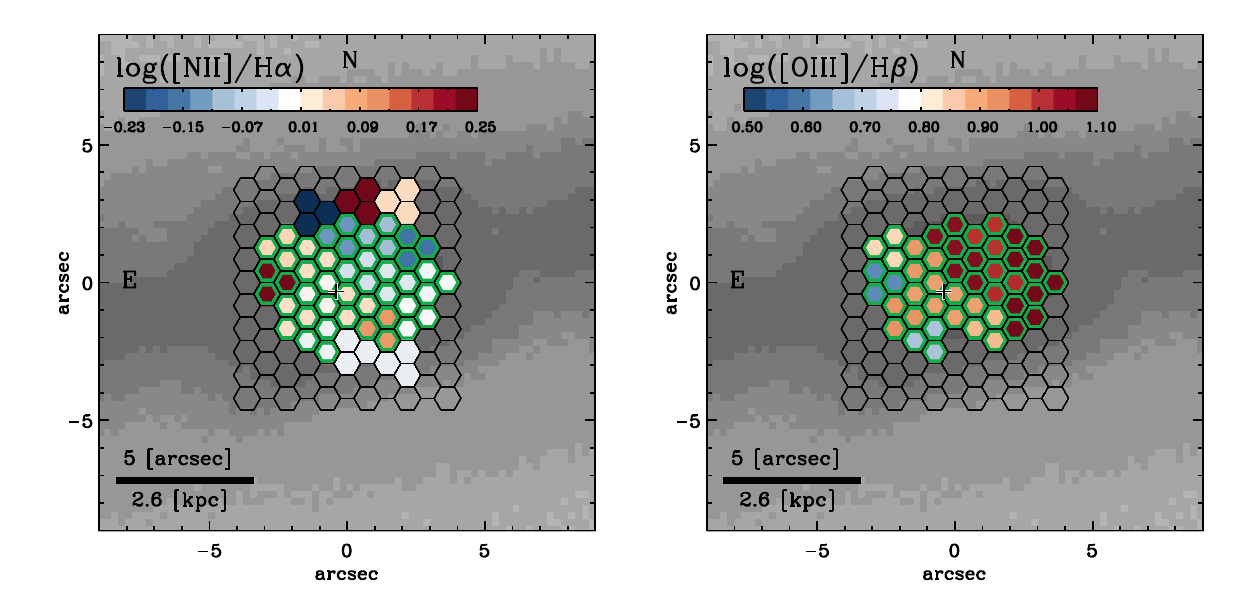}
    \caption{Spatially resolved emission-line ratio maps of NGC~6552 derived from KOOLS-IFU observations. 
Left: $\log([\mathrm{N\,II}]/\mathrm{H}\alpha)$, shown for bins with A/N$([\mathrm{N\,II}])>3$ and A/N$(\mathrm{H}\alpha)>3$. 
Right: $\log([\mathrm{O\,III}]/\mathrm{H}\beta)$, shown for bins with A/N$([\mathrm{O\,III}])>3$ and A/N$(\mathrm{H}\beta)>3$. 
Green thick outlines mark bins classified as Seyfert on the BPT diagram, 
requiring all four emission lines (\NII, \Ha, \OIII, and \Hb) to satisfy A/N$>3$. 
The smaller spatial extent of the Seyfert-classified region in the right panel reflects the lower S/N of H$\beta$ 
rather than a physical change in ionization state. 
    \label{fig:line_ratio_maps}}
\end{figure*}

\subsection{Kiloparsec-scale \OIII\ Kinematics}\label{sec:results_kinematics}

\subsubsection{Two-dimensional kinematic maps}\label{sec:results_kin_maps}

Figure~\ref{fig:OIII_kinematics} summarizes the spatially resolved \OIII\ kinematics using non-parametric metrics. 
The $W_{80}$ map (Figure~\ref{fig:OIII_kinematics}a) shows widespread line broadening across the inner region, 
with $W_{80}$ spanning $\sim530$--$830$~km~s$^{-1}$. 
Elevated widths ($W_{80}\gtrsim600$~km~s$^{-1}$, adopting the conservative threshold of \citealt{Harrison14}) extend to projected radii of order $\sim1.5$ kpc, 
indicating kpc-scale kinematic disturbance in the ionized gas.

The median-velocity field $v_{50}$ (Figure~\ref{fig:OIII_kinematics}b) exhibits modest line-of-sight velocities relative to 
systemic, typically within $|v_{50}|\lesssim100$~km~s$^{-1}$. 
The velocity asymmetry parameter $\Delta v$ (Figure~\ref{fig:OIII_kinematics}c) reaches values from 
$\sim -170$ to $\sim +60$~km~s$^{-1}$, implying spatially varying dominance of blue vs.\ red wings in the \OIII\ profile, 
with blue-wing asymmetry predominating across most of the field. 
The outflow A/N map (Figure~\ref{fig:OIII_kinematics}d) shows that the shifted component is detected across much of the field, 
although the spatial distribution of outflow A/N does not simply mirror the $W_{80}$ pattern, 
reflecting the distinct sensitivities of the non-parametric and parametric measures.

\subsubsection{Radial behavior of $W_{80}$ and $\log([\mathrm{O\,III}]/\mathrm{H}\beta)$}\label{sec:results_w80_radial}

Figure~\ref{fig:W80_vs_radius}a shows $W_{80}$ as a function of projected galactocentric distance $r$. 
The profile declines monotonically from the nucleus outward: a weighted linear fit, 
$W_{80} = a + br$, yields $a = 803 \pm 11$~km~s$^{-1}$ and 
$b = -83 \pm 10$~km~s$^{-1}$~kpc$^{-1}$ (solid line in Figure~\ref{fig:W80_vs_radius}a). 
We also tested a quadratic model, $W_{80} = a + br + cr^{2}$, 
but the additional term is not statistically justified: the $F$-test yields $p = 0.80$, 
confirming that the linear model is sufficient. 
The reduced chi-square of the linear fit is $\chi^{2}_{\nu} \approx 7.4$, 
indicating substantial intrinsic scatter around the smooth trend. 
Over the KOOLS-IFU field, $W_{80}$ spans $\sim$530--830~km~s$^{-1}$ (Figure~\ref{fig:OIII_kinematics}a), 
with the highest values near the nucleus and a systematic decline at larger radii (Figure~\ref{fig:W80_vs_radius}a). 
The physical interpretation of this centrally peaked profile is discussed in Section~\ref{sec:W80_origin}.

Figure~\ref{fig:W80_vs_radius}b shows the same radial dependence for $\log([\mathrm{O\,III}]/\mathrm{H}\beta)$, 
computed for the $N=15$ bins satisfying A/N $> 3$ in both \OIII\ and \Hb. 
This diagnostic also exhibits a statistically significant negative correlation with $r$ (Spearman $\rho = -0.63$, $p = 0.014$; Table~\ref{tab:corr_tests}), 
demonstrating that both the kinematic broadening and the excitation state vary systematically with projected distance from the nucleus. 
The implications of this shared radial dependence, together with the absence of a direct $W_{80}$--$\log([\mathrm{O\,III}]/\mathrm{H}\beta)$ correlation at fixed $r$ 
(Section~\ref{sec:results_coupling}), are discussed in Section~\ref{sec:decoupling}.

\begin{figure}
\centering
\includegraphics[width=\columnwidth]{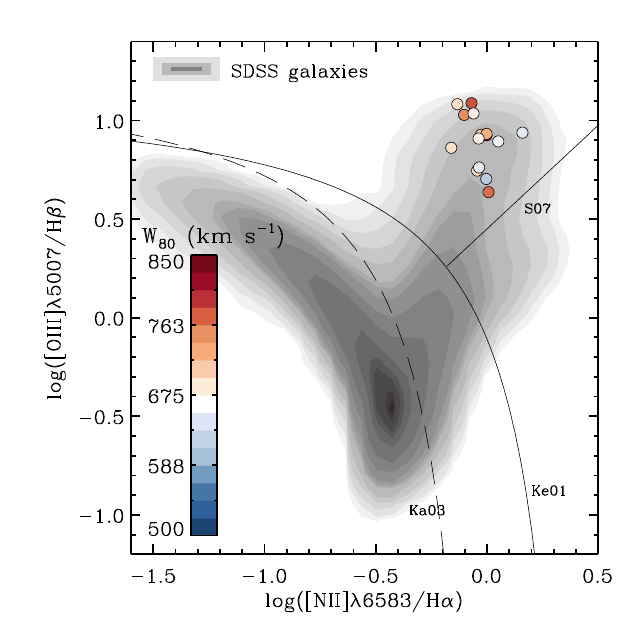}
\caption{
$[\mathrm{O\,III}]\lambda5007/\mathrm{H}\beta$ versus $[\mathrm{N\,II}]\lambda6583/\mathrm{H}\alpha$ BPT diagram 
for the $N_{\rm bins}=15$ spatial bins satisfying A/N$>3$ in all four emission lines 
(\NII, \Ha, \OIII, \Hb). 
Data points are color-coded by the non-parametric velocity width $W_{80}$ (km\,s$^{-1}$), 
as indicated by the color bar. 
Error bars represent $1\sigma$ uncertainties propagated from the emission-line flux errors, 
and in all cases the uncertainties are smaller than the symbol size. 
The background grey contours show the distribution of $\sim$200,000 SDSS emission-line galaxies 
measured by the OSSY database \citep{Oh11} for reference. 
The dashed curve shows the empirical star-forming/AGN demarcation of \citet{Kauffmann03}, 
and the solid curve shows the theoretical maximum starburst line of \citet{Kewley01}. 
The solid straight line is the empirical demarcation between Seyfert and LINER \citep{Schawinski07}. 
All bins lie in the Seyfert region above the lines of \citet{Kewley01} and \citet{Schawinski07}. 
No systematic segregation of high- and low-$W_{80}$ bins is apparent in BPT space, 
consistent with the null correlation reported in Table~\ref{tab:corr_tests} and Figure~\ref{fig:W80_vs_OIIIHb}.
}
\label{fig:bpt}
\end{figure}

\begin{figure*}
    \centering
    \includegraphics[width=0.95\linewidth]{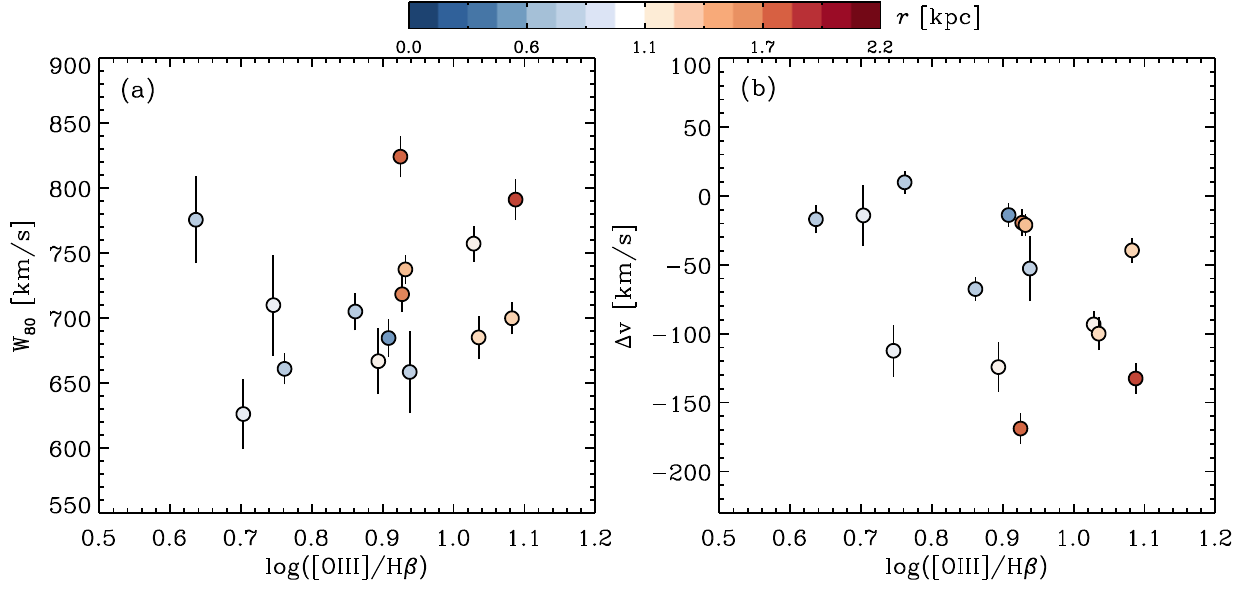}
\caption{(a) Non-parametric velocity width $W_{80}$ and
(b) velocity asymmetry parameter $\Delta v = (v_{05}+v_{95})/2 - v_{50}$ 
of the \OIII\ emission line as a function of $\log([\mathrm{O\,III}]/\mathrm{H}\beta)$ 
for spatial bins satisfying A/N$>3$ in both lines ($N_{\rm bins}=15$). 
Data points are color-coded by projected galactocentric distance $r$ (kpc) from the JWST F560W-defined nucleus, 
as indicated by the color bar.
Vertical error bars represent Monte Carlo uncertainties. 
$W_{80}$ shows no significant correlation with $\log([\mathrm{O\,III}]/\mathrm{H}\beta)$ (Spearman $\rho=+0.24$, $p=0.41$).
$\Delta v$ likewise shows no significant correlation (Spearman $\rho=-0.40$, $p=0.14$; Pearson $r=-0.39$, $p=0.15$).
In panel~(a), $W_{80}$ shows a significant dependence on projected radius (Spearman $\rho=-0.54$, $p=0.039$)
but not with the excitation ratio, consistent with the radially driven kinematic structure discussed in Section~\ref{sec:W80_origin}. 
Correlation statistics are summarized in Table~\ref{tab:corr_tests}.}
\label{fig:W80_vs_OIIIHb}
\end{figure*}

\subsection{Spatially Resolved Excitation Diagnostics}\label{sec:results_excitation}
Figure~\ref{fig:line_ratio_maps} presents two excitation-sensitive line-ratio maps:
$\log([\mathrm{N\,II}]\lambda6583/\mathrm{H}\alpha)$ and $\log([\mathrm{O\,III}]\lambda5007/\mathrm{H}\beta)$. 
Both maps are constructed using only spatial bins that satisfy A/N$>3$ for the relevant lines, 
and therefore the spatial footprint differs between the two ratios. 
Within the region sampled, $\log([\mathrm{N\,II}]/\mathrm{H}\alpha)$ shows patchy spatial structure with no clear monotonic trend, 
while $\log([\mathrm{O\,III}]/\mathrm{H}\beta)$ exhibits a systematic increase from east to west across the field. 
Whether this gradient reflects a true spatial variation in ionization conditions or is driven by differential extinction across the field cannot be determined from the present data alone. 
We also outline bins that can be classified on the BPT diagram (i.e., bins with A/N$>3$ in all four lines). 
The Seyfert-classified bins are concentrated near the nucleus (Figure~\ref{fig:line_ratio_maps}).

Figure~\ref{fig:bpt} shows the $[\mathrm{O\,III}]\lambda5007/\mathrm{H}\beta$ versus 
$[\mathrm{N\,II}]\lambda6583/\mathrm{H}\alpha$ BPT diagram for the $N_{\rm bins}=15$ bins satisfying A/N$>3$ in all four emission lines. 
The data points are color-coded by $W_{80}$ to visualize any correspondence between 
kinematic broadening and the position in excitation-diagnostic space. 
The empirical demarcation of \citet{Kauffmann03} and the theoretical maximum starburst 
line of \citet{Kewley01} are overplotted. 
All $N_{\rm bins}=15$ bins fall in the Seyfert region, above the lines of \citet{Kewley01} and \citet{Schawinski07}, 
confirming that AGN photoionization dominates the line emission across the sampled area. 
The bins span a relatively narrow range in $\log([\mathrm{O\,III}]/\mathrm{H}\beta)$ ($\approx 0.6$--$1.1$), 
and no systematic color gradient is apparent in the diagram. 
Bins with high and low $W_{80}$ are interspersed rather than segregated in BPT space, 
a visual impression that we quantify in Section~\ref{sec:results_coupling}.

Since \Hb\ is the limiting line in many bins, 
the spatial extent of BPT-classified bins is primarily driven by the \Hb\ S/N rather than necessarily reflecting a sharp physical boundary in excitation. 
A visual comparison between the excitation maps (Figure~\ref{fig:line_ratio_maps}) and the kinematic maps (Figure~\ref{fig:OIII_kinematics}) does not reveal 
an obvious one-to-one correspondence between the highest $W_{80}$ regions and the extrema of $\log([\mathrm{O\,III}]/\mathrm{H}\beta)$.

\subsection{Testing the Kinematics--Excitation Coupling}\label{sec:results_coupling}
We test for a resolved coupling between kinematic measures 
and the excitation-sensitive ratio $\log([\mathrm{O\,III}]/\mathrm{H}\beta)$ 
using only spatial bins with robust measurements of both \OIII\ and H$\beta$ (A/N$>3$ in both lines; $N_{\rm bins}=15$). 
The correlation and significance tests are summarized in Table~\ref{tab:corr_tests}. 

Figure~\ref{fig:W80_vs_OIIIHb} shows the $W_{80}$--$\log([\mathrm{O\,III}]/\mathrm{H}\beta)$ distribution. 
We find no statistically significant correlation between $W_{80}$ and $\log([\mathrm{O\,III}]/\mathrm{H}\beta)$ in 
either rank-based or linear tests (Spearman $\rho = +0.24$, $p = 0.41$; Pearson $r = +0.22$, $p = 0.40$; Table~\ref{tab:corr_tests}). 
This conclusion remains unchanged when controlling for projected radius 
via a partial correlation test ($\rho_{\rm partial} = -0.17$, $p = 0.56$), 
and when performing a residual--residual test after subtracting quadratic radial trends from both quantities ($\rho = -0.046$, $p = 0.87$).

The velocity asymmetry parameter $\Delta v$ likewise shows no significant correlation with $\log([\mathrm{O\,III}]/\mathrm{H}\beta)$ 
(Spearman $\rho = -0.40$, $p = 0.14$; Pearson $r = -0.39$, $p = 0.15$; Table~\ref{tab:corr_tests}). 
The Monte Carlo uncertainties on $\Delta v$ (Section~\ref{sec:obs_nonparam}) have a median of $\sim 11$~km~s$^{-1}$, 
well below the observed bin-to-bin dispersion ($\sigma \approx 54$~km~s$^{-1}$), 
confirming that measurement errors do not dominate the scatter. 

$W_{80}$ itself shows a significant dependence on projected radius for the bins used in the comparison 
(Spearman $\rho = -0.54$, $p = 0.039$), 
and $\log([\mathrm{O\,III}]/\mathrm{H}\beta)$ also shows a significant radial decline 
($\rho = -0.63$, $p = 0.014$). 
Despite these shared radial trends, the direct $W_{80}$--$\log([\mathrm{O\,III}]/\mathrm{H}\beta)$ 
correlation is not significant, and controlling for projected radius yields 
$\rho_{\rm partial} = -0.17$ ($p = 0.56$), 
confirming that the covariation is driven by the shared radial dependence 
rather than by a physical coupling between kinematics and excitation. 
The limited sample size ($N_{\rm bins} = 15$) restricts our ability to detect moderate correlations 
($|\rho| \sim 0.3$--$0.5$; Section~\ref{sec:limitations}), 
but the absence of significant trends for both $W_{80}$ and $\Delta v$ supports 
a consistent picture of kinematic--excitation decoupling. 
The implications of these null results are discussed in Section~\ref{sec:decoupling}.

\begin{deluxetable}{lccc}
\tablecaption{Correlation tests between $W_{80}$, $\Delta v$, 
excitation diagnostics, and projected radius
\label{tab:corr_tests}}
\tablewidth{0pt}
\tablehead{
\colhead{Relation} & \colhead{Test} & \colhead{Statistic} & \colhead{$p$-value}
}
\startdata
\multicolumn{4}{c}{\OIIIHb\ diagnostics ($N_{\rm bins} = 15$)} \\
\hline
$W_{80}$ vs.\ $\log([\mathrm{O\,III}]/\mathrm{H}\beta)$ & Spearman & $\rho = +0.24$ & 0.41 \\
$W_{80}$ vs.\ $\log([\mathrm{O\,III}]/\mathrm{H}\beta)$ & Pearson & $r = +0.22$ & 0.40 \\
$W_{80}$ vs.\ $\log([\mathrm{O\,III}]/\mathrm{H}\beta)$ (control $r$) & Partial Spearman & $\rho_{\rm partial} = -0.17$ & 0.56 \\
$W_{80,\rm resid}$ vs.\ $\log([\mathrm{O\,III}]/\mathrm{H}\beta)_{\rm resid}$ & Spearman & $\rho = -0.046$ & 0.87 \\
$\Delta v$ vs.\ $\log([\mathrm{O\,III}]/\mathrm{H}\beta)$ & Spearman & $\rho = -0.40$ & 0.14 \\
$\Delta v$ vs.\ $\log([\mathrm{O\,III}]/\mathrm{H}\beta)$ & Pearson & $r = -0.39$ & 0.15 \\
$\Delta v$ vs.\ $\log([\mathrm{O\,III}]/\mathrm{H}\beta)$ (control $r$) & Partial Spearman & $\rho_{\rm partial} = -0.035$ & 0.91 \\
$W_{80}$ vs.\ $r$ & Spearman & $\rho = -0.54$ & 0.039 \\
$\log([\mathrm{O\,III}]/\mathrm{H}\beta)$ vs.\ $r$ & Spearman & $\rho = -0.63$ & 0.014 \\
\hline
\multicolumn{4}{c}{\SIIHa\ diagnostics ($N_{\rm bins} = 12$)} \\
\hline
$W_{80}$ vs.\ $\log([\mathrm{S\,II}]/\mathrm{H}\alpha)$ & Spearman & $\rho = -0.25$ & 0.43 \\
$W_{80}$ vs.\ $\log([\mathrm{S\,II}]/\mathrm{H}\alpha)$ & Pearson & $r = -0.17$ & 0.60 \\
$W_{80}$ vs.\ $\log([\mathrm{S\,II}]/\mathrm{H}\alpha)$ (control $r$) & Partial Spearman & $\rho_{\rm partial} = -0.24$ & 0.46 \\
$\Delta v$ vs.\ $\log([\mathrm{S\,II}]/\mathrm{H}\alpha)$ & Spearman & $\rho = -0.15$ & 0.63 \\
$\Delta v$ vs.\ $\log([\mathrm{S\,II}]/\mathrm{H}\alpha)$ & Pearson & $r = -0.23$ & 0.45 \\
$\Delta v$ vs.\ $\log([\mathrm{S\,II}]/\mathrm{H}\alpha)$ (control $r$) & Partial Spearman & $\rho_{\rm partial} = -0.47$ & 0.12 \\
$\log([\mathrm{S\,II}]/\mathrm{H}\alpha)$ vs.\ $r$ & Spearman & $\rho = +0.098$ & 0.76 \\
\enddata
\tablecomments{The \OIIIHb\ tests (upper section) use bins with A/N$>3$ 
in both \OIII\ and \Hb\ ($N_{\rm bins} = 15$, Figure~\ref{fig:W80_vs_OIIIHb}). 
The \SIIHa\ tests (lower section) use bins with A/N$>3$ in both 
$[\mathrm{S\,II}]\lambda6717$ and $[\mathrm{S\,II}]\lambda6731$ ($N_{\rm bins} = 12$). 
All $p$-values are permutation-based (two-sided). 
Residuals are computed after subtracting quadratic radial trends from both quantities. 
The velocity asymmetry parameter is defined as $\Delta v = (v_{05}+v_{95})/2 - v_{50}$.}
\end{deluxetable}

\section{Discussion}\label{sec:discussion}

\subsection{The Centrally Peaked $W_{80}$ Profile}\label{sec:W80_origin}

The $W_{80}$ radial profile declines monotonically from the nucleus outward (Figure~\ref{fig:W80_vs_radius}), 
with the highest values ($\sim$830~km~s$^{-1}$) at the smallest projected radii. 
This centrally peaked pattern is consistent with the majority of IFU surveys of AGN host galaxies, 
which find that $W_{80}$ is maximized at the nucleus and decreases with increasing distance 
\citep[e.g.,][]{Liu13, McElroy15, RuschelDutra21, Alban24}. 
The monotonic decline is naturally produced when the strongest kinematic disturbance originates near the AGN and 
diminishes outward due to deceleration, momentum loss through entrainment of ambient gas, 
decreasing surface brightness of the broad outflow component, 
and geometric dilution in a biconical or wide-angle outflow geometry.

A purely rotational origin for the observed $W_{80}$ values is disfavored by the moderate amplitude of the median-velocity field 
($|v_{50}| \lesssim 100$~km~s$^{-1}$; Figure~\ref{fig:OIII_kinematics}) compared to the large line widths 
($W_{80} \sim 530$--$830$~km~s$^{-1}$). 
While beam smearing can broaden emission lines where the velocity gradient is steep, 
our forward model of a pure rotating disk ($v_{\rm rot,max} = 100$~km~s$^{-1}$) convolved with the observed PSF (FWHM $= 1.3''$) 
produces a maximum beam-smeared $W_{80}$ of only $\approx 262$~km~s$^{-1}$ at the nucleus, 
decreasing monotonically outward. 
This confirms that beam smearing alone cannot account for the observed $W_{80} \sim 530$--$830$~km~s$^{-1}$ 
and that the radial $W_{80}$ profile is dominated by astrophysical broadening.
Even in the outermost bins, the observed $W_{80}$ values remain well above the $\sim 200$--$300$~km~s$^{-1}$ range 
characteristic of quiescent disk kinematics in nearby Seyfert galaxies \citep{RuschelDutra21}, 
reinforcing that the elevated line widths across the entire field cannot be attributed to instrumental broadening.

The intrinsic scatter around the linear fit is substantial ($\chi^{2}_{\nu} \approx 7.4$; Figure~\ref{fig:W80_vs_radius}), 
implying that the ionized-gas kinematics are not purely radial and likely vary azimuthally due to clumpy ISM structure, 
asymmetric interaction sites, or multiple kinematic components within individual spatial bins. 
This is consistent with the spatially localized regions of high outflow A/N (Figure~\ref{fig:OIII_kinematics}), 
supporting a picture in which a centrally launched outflow interacts with an inhomogeneous host ISM. 
Such clumpy ionized-gas morphologies are commonly observed in nearby Seyfert galaxies 
with VLT/MUSE at higher spatial resolution \citep{Venturi18,Marconcini23,Marconcini25}. 
In confirmed AGN outflow hosts such as Circinus and NGC~4945, line-of-sight superposition of disk and outflow components 
produces $W_{80} \sim 600$--$800$~km~s$^{-1}$ through the combination of moderate velocities and velocity dispersions \citep{Marconcini23,Marconcini25}, 
consistent with our finding that $W_{80}$ is governed by spatial dynamical structure rather than the local excitation state.

The observed $W_{80}$ range of $\sim$530--830~km~s$^{-1}$ in NGC~6552 is moderate compared to more luminous AGN samples. 
\citet{Harrison14} reported $\langle W_{80}\rangle\approx670$--$1450$~km~s$^{-1}$ for 
luminous type~2 AGN observed with GMOS-IFU, 
and \citet{Liu13} measured $W_{80}\approx490$--$1850$~km~s$^{-1}$ for 
higher-redshift ($z\sim0.55$) quasars. 
Among the 22 $z<0.1$ X-ray AGN studied with MUSE by \citet{Kakkad22}, 
integrated $W_{80}$ values range from $\sim290$ to $\sim1450$~km~s$^{-1}$, 
with about three-quarters of the sample exceeding $500$~km~s$^{-1}$. 
NGC~6552 falls well within the outflow-dominated portion of this distribution. 
The AGNIFS GMOS-IFU survey of 30 $z\leq0.02$ hosts similarly detected 
outflows in 21 nuclei at 50--300~pc resolution using $W_{80}$ as the primary kinematic tracer \citep{RuschelDutra21}. 
Using the same KOOLS-IFU instrument and non-parametric methodology, 
\citet{Oh25} measured spatially resolved $W_{80} \approx 310$--$860$~km~s$^{-1}$ 
in the nearby Seyfert galaxy Mrk~766, comparable to the range found here for NGC~6552. 
The monotonically declining $W_{80}$ profile in NGC~6552 is consistent with the centrally peaked profiles reported in these surveys.

Our optical results complement the mid-infrared integral-field findings of \citet{AlvarezMarquez23}, 
who reported the first clear evidence for a nuclear outflow in NGC~6552 using JWST/MIRI MRS observations. 
In their analysis, the outflow was detected as a blue-shifted velocity component in high-excitation and coronal emission lines 
spanning ionization potentials of 27--187~eV, with an average peak velocity offset of $-127 \pm 45$~km~s$^{-1}$ 
and a maximal outflow velocity of $698 \pm 80$~km~s$^{-1}$. 
This mid-infrared outflow was spatially unresolved, confined to $< 0.2$~kpc from the nucleus. 
In contrast, our KOOLS-IFU \OIII\ observations reveal that the kinematic disturbance extends well beyond this nuclear region: 
$W_{80}$ is maximized near the nucleus ($\sim 830$~km~s$^{-1}$) and declines monotonically outward, 
yet remains elevated ($\gtrsim 600$~km~s$^{-1}$) out to projected radii of $\sim$1.5~kpc 
before falling to $\sim$530--590~km~s$^{-1}$ at the field edges. 
This monotonically declining profile suggests that the nuclear outflow launched within $< 0.2$~kpc 
gradually decelerates as it propagates outward through the host ISM.

The two datasets also highlight a clear phase-dependent behavior. 
\citet{AlvarezMarquez23} found that the warm molecular H$_{2}$ lines in the nuclear spectrum are symmetric and 
show no evidence of outflowing material (FWHM $= 312\pm34$~km~s$^{-1}$, 
consistent with the systemic velocity component), 
whereas the highly ionised atomic lines exhibit prominent blue-shifted wings. 
Our optical \OIII\ emission traces the lower-density, less-obscured ionised gas at larger radii. 
The persistence of elevated $W_{80}$ values ($\gtrsim 600$~km~s$^{-1}$) out to $r \sim 1.5$~kpc indicates that 
the fully ionised nuclear outflow retains sufficient velocity to produce broad optical line profiles well beyond the spatially unresolved mid-infrared detection zone. 
A joint kinematic comparison using matched apertures and consistent non-parametric metrics across the mid-infrared and optical 
would provide a powerful constraint on the radial velocity and energy budget of the multi-phase outflow in this Compton-thick AGN.

\subsection{Decoupling Between Kinematics and Ionization}\label{sec:decoupling}

Despite the pronounced radial structure in $W_{80}$, 
we find no statistically significant correlation between $W_{80}$ and 
the ionization-sensitive ratio \OIIIHb\ across bins with reliable measurements (Figure~\ref{fig:W80_vs_OIIIHb}). 
Both rank-based and linear tests yield null results (Spearman $\rho=+0.24$, $p=0.41$; Pearson $r=+0.22$, $p=0.40$), 
and the conclusion remains unchanged when controlling for projected radius via a partial correlation analysis 
($\rho_{\rm partial}=-0.17$, $p=0.56$) or 
when correlating residuals after subtracting radial trends ($\rho=-0.046$, $p=0.87$). 
Both $W_{80}$ and $\log([\mathrm{O\,III}]/\mathrm{H}\beta)$ decline significantly with projected radius 
(Spearman $\rho = -0.54$, $p = 0.039$ and $\rho = -0.63$, $p = 0.014$, respectively), 
yet no direct correlation between the two quantities survives after controlling for radius 
($\rho_{\rm partial} = -0.17$, $p = 0.56$). 
This indicates that the radial gradients in kinematics and excitation are independently driven, 
likely reflecting the deceleration of the outflow and 
the dilution of the AGN radiation field at larger radii, respectively, 
rather than a causal link between the two.

The absence of a $W_{80}$--\OIIIHb\ correlation is broadly consistent with the three-dimensional diagnostic framework of \citet{DAgostino19a}, 
who showed that IFU data of a nearby galaxy separate into two distinct mixing sequences. 
The first is a star-formation--AGN sequence in which velocity dispersion shows little systematic change with increasing emission-line ratios. 
The second is a star-formation--shock sequence in which the two quantities are tightly correlated \citep[see also][]{DAgostino19b}. 
Although their framework was developed for spaxels spanning a range of excitation mechanisms, 
the underlying physical picture applies here. 
When AGN photoionization dominates (as indicated by the uniformly Seyfert-classified bins in NGC~6552; Figure~\ref{fig:bpt}), 
elevated line ratios reflect the hardness of the AGN radiation field rather than shock-enhanced kinematics, 
and velocity broadening can vary independently of \OIIIHb. 
We note that $W_{80}$ (a non-parametric measure of the full line profile including multi-component wings) differs 
from the single-component velocity dispersion used by \citet{DAgostino19a}, 
but the qualitative expectation of decoupled kinematics and excitation in AGN-dominated environments is the same.

The velocity asymmetry parameter $\Delta v$ also shows no significant 
$\Delta v$--$\log([\mathrm{O\,III}]/\mathrm{H}\beta)$ correlation 
(Spearman $\rho = -0.40$, $p = 0.14$), 
and this remains unchanged when controlling for projected radius 
($\rho_{\rm partial} = -0.035$, $p = 0.91$).
The absence of a significant correlation for both $W_{80}$ and $\Delta v$ indicates that 
in this AGN-dominated environment neither the total line width nor the profile asymmetry is coupled to the local excitation ratio. 
This result strengthens the interpretation that the observed line broadening and asymmetry are governed by dynamical structure, 
projection effects, and line-of-sight superposition of multiple kinematic components rather than by excitation-driven processes.

Several physical and observational effects can plausibly produce the observed absence of any kinematic--excitation coupling. 
First, $W_{80}$ is highly sensitive to faint high-velocity wings. 
If each spatial bin contains a mixture of a bright narrow component (e.g., AGN-photoionized or ambient disk gas) 
plus a fainter broad component (e.g., outflowing or shocked gas), 
the line ratio can be dominated by the narrow component while $W_{80}$ is boosted by the broad wings. 
In this ``mixing'' scenario, substantial kinematic disturbance can coexist with only modest changes in \OIIIHb, 
whereas $\Delta v$ --- being sensitive to the relative strength rather than absolute breadth of the asymmetric wing --- 
might in principle track variations in the outflow contribution, although no significant correlation is detected in our data. 

Second, shock excitation does not necessarily translate into a monotonic increase in \OIIIHb. 
Depending on shock velocity, pre-shock density, magnetic parameter, and metallicity, 
shocks can enhance low-ionization lines (e.g., \NII, \SII, \OI) more strongly than \OIII, 
and therefore a single ratio may be an incomplete tracer of the shock fraction. 
In NGC~6552, the combination of a compact Seyfert-like nucleus and extended line emission, 
together with the H$\beta$-limited BPT classification footprint (Figure~\ref{fig:line_ratio_maps}), 
suggests that multiple excitation sources (AGN radiation, shocks, and/or diffuse ionization) may coexist spatially.
To test whether a shock-sensitive diagnostic yields a different result, 
we repeated the correlation analysis using $\log([\mathrm{S\,II}]\lambda\lambda6717{,}6731/\mathrm{H}\alpha)$ for the $N_{\rm bins} = 12$ bins in which 
both $[\mathrm{S\,II}]$ lines are detected with A/N $> 3$ (see Appendix~\ref{app:specfit_examples} for representative spectral fitting quality in the \Ha\ region). 
Neither $W_{80}$ nor $\Delta v$ shows a significant correlation with $\log([\mathrm{S\,II}]/\mathrm{H}\alpha)$ 
(Spearman $\rho = -0.25$, $p = 0.43$ and $\rho = -0.15$, $p = 0.63$, respectively; Table~\ref{tab:corr_tests}), 
and this conclusion is unchanged when controlling for projected radius 
($\rho_{\rm partial} = -0.24$, $p = 0.46$ for $W_{80}$, 
and $\rho_{\rm partial} = -0.47$, $p = 0.12$ for $\Delta v$). 
Unlike the \OIIIHb\ ratio, $\log([\mathrm{S\,II}]/\mathrm{H}\alpha)$ 
does not show a significant radial trend in this sample ($\rho = +0.098$, $p = 0.76$). 
The concordance of null results across both high-ionization (\OIIIHb) 
and low-ionization (\SIIHa) diagnostics reinforces the conclusion that 
the kinematic--excitation decoupling in NGC~6552 is not an artifact of using a single line ratio. 
A more complete assessment would benefit from multi-line mixing sequences and explicit two-component spectral decomposition.

Third, the ratio \OIIIHb\ is sensitive to gas density and metallicity, but not significantly to differential extinction. 
Because the two lines are separated by only $\sim$146~\AA, 
the maximum \citet{Calzetti00} differential effect amounts to $\sim$0.03~dex for $E(B-V)_{\rm int}=0.5$, 
less than 6 percent of the observed $\log([\mathrm{O\,III}]/\mathrm{H}\beta)$ dynamic range across our spatial bins. 
The observed Balmer decrements in our spatial bins range from 
$\mathrm{H}\alpha/\mathrm{H}\beta \approx 2.9$ (consistent with Case~B recombination) 
to $\approx 5.14$, indicating spatially non-uniform internal extinction 
with $E(B-V)_{\rm int}$ varying from $\approx 0$ to $\approx 0.5$~mag across the field. 
However, as quantified above, the resulting differential effect on the observed $\log([\mathrm{O\,III}]/\mathrm{H}\beta)$ is small.
In our data, the smaller spatial extent of the Seyfert-classified region in the \OIIIHb\ map 
is driven by the lower S/N of H$\beta$ (Figure~\ref{fig:line_ratio_maps}), 
which can further dilute subtle trends and reduce the dynamic range of the ratio.

Overall, the absence of a resolved correlation between either $W_{80}$ or $\Delta v$ and $\log([\mathrm{O\,III}]/\mathrm{H}\beta)$ 
indicates that in this hard X-ray selected obscured AGN, 
both the total line broadening and the line-profile asymmetry are governed primarily by 
spatial dynamical structure and line-of-sight superposition of multiple kinematic components. 
In an environment where AGN photoionization uniformly dominates the excitation 
(all bins classified as Seyfert; Figure~\ref{fig:bpt}), 
variations in $W_{80}$ and $\Delta v$ reflect differences in outflow geometry, ISM clumpiness, 
and the relative contribution of narrow and broad kinematic components along the line of sight, 
rather than local changes in ionization conditions.
A robustness check based on the Bayesian Information Criterion of the two-component \OIII\ fit (Appendix~\ref{app:appendix_bic}) confirms that 
the $\Delta v$--$\log([\mathrm{O\,III}]/\mathrm{H}\beta)$ decoupling is preserved across BIC-based subsamples 
once the shared radial dependence is controlled for, while a suggestive $W_{80}$--$\log([\mathrm{O\,III}]/\mathrm{H}\beta)$ coupling that emerges 
in the most stringent BIC subsample is small in scale ($N_{\mathrm{bins}} = 7$) and warrants further investigation.

\subsection{Order-of-Magnitude Outflow Energetics}\label{sec:energetics}

Although deriving precise outflow energetics is beyond the scope of this work, 
we provide order-of-magnitude estimates to place NGC~6552 in the context of ionized outflow properties in local AGNs. 
Following the formalism of \citet{CanoDiaz12} as applied by \citet{Oh25}, 
we estimate the ionized gas mass, mass outflow rate, and kinetic power from the spatially integrated \OIII\ outflow-component luminosity. 
We sum the flux of the shifted \OIII\ component over all spatial bins where it is detected with A/N $> 2$. 
Bins below this threshold are excluded because the shifted component is not reliably measured, 
so the resulting luminosity should be regarded as a lower limit. 
After applying a dust extinction correction to each bin using the observed Balmer decrement (\Ha/\Hb) following \citet{Cardelli89}, 
assuming an intrinsic ratio of $\mathrm{H}\alpha/\mathrm{H}\beta = 2.86$ (Case~B recombination at $T = 10^{4}$~K; \citealt{Osterbrock06}), 
the extinction-corrected outflow luminosity is $\log(L_{[\mathrm{O\,III}],\rm out}/\mathrm{erg\,s}^{-1}) \approx 41.1$.

We characterize the outflow velocity from the shifted (outflow) \OIII\ component rather than from the non-parametric width $W_{80}$, 
so as to use a quantity that traces the kinematics of the outflowing gas itself. 
For each bin with a detected shifted component (A/N $> 2$), we adopt a maximal outflow velocity 
$v_{\rm out} = |v_{\rm shift}| + 2\sigma_{\rm shift}$, 
where $v_{\rm shift}$ is the velocity offset of the shifted component relative to the systemic (stellar) velocity 
and $\sigma_{\rm shift}$ is its velocity dispersion. 
This corresponds to the maximal-velocity convention $v_{98} = v_{50} - 2\sigma$ of \citet{Rupke13}, 
and is closely related to the broad-component velocity $v_{\rm out} = \mathrm{FWHM}_{\rm broad}/2 + |v_{\rm broad} - v_{\rm narrow}|$ adopted by \citet{Fluetsch19}. 
The flux-weighted mean over these bins is $v_{\rm out} \approx 900$~km~s$^{-1}$, 
which we adopt together with $R_{\rm out} = 1.0$~kpc as a representative outflow radius, 
consistent with the scale at which the outflow is robustly detected (Figure~\ref{fig:OIII_kinematics}d). 
This velocity agrees with the maximum observed $W_{80}$ to within $\sim 10\%$, 
so that the choice of velocity definition propagates as only $\approx 0.1$~dex into $\dot{E}_{K}$, 
well within the $\sim 1.3$~dex systematic uncertainty introduced by the unconstrained electron density.
To account for this dominant uncertainty, and following \citet{Davies20}, 
who demonstrated that  $[\mbox{S\,{\sc ii}}]$-based density estimates can underestimate 
the true outflow density by one to two orders of magnitude, 
we bound the results by assuming $n_{e} = 50$--$1000$~cm$^{-3}$. 
Because $M_{\rm out} \propto n_{e}^{-1}$, this assumed range propagates directly into all derived quantities, 
producing a factor of $\sim 20$ ($\approx 1.3$~dex) uncertainty that dominates over other error sources. 
The ionized outflow mass ranges from $\log(M_{\rm out}/M_{\odot}) \approx 4.8$ ($n_{e} = 1000$~cm$^{-3}$) to 
$\approx 6.1$ ($n_{e} = 50$~cm$^{-3}$). 
The mass outflow rate ranges from $\dot{M}_{\rm out} \approx 0.2$~$M_{\odot}$~yr$^{-1}$ to $\approx 3.6$~$M_{\odot}$~yr$^{-1}$. 
The kinetic coupling efficiency is $\dot{E}_{K}/L_{\rm bol} \approx 0.01\%$--$0.28\%$, 
well below the $\sim 5\%$ threshold invoked by energy-driven feedback models \citep{King10, Zubovas12}. 
This low efficiency is consistent with the general finding that  $[\mbox{O\,{\sc iii}}]$-based estimates trace only 
the ionized phase of a multi-phase outflow and therefore represent a lower limit on the total outflow budget. 
Using $L_{\rm bol} = 10^{44.51}$~erg~s$^{-1}$ \citep{Koss22_catalog_and_data} and a radiative efficiency $\eta = 0.1$, 
the mass accretion rate is $\dot{m}_{\rm accr} \approx 0.06$~$M_{\odot}$~yr$^{-1}$, 
yielding a mass loading factor $\dot{M}_{\rm out}/\dot{m}_{\rm accr} \approx 3$--$62$, again spanning the assumed density range.

These estimates carry systematic uncertainties of at least an order of magnitude, 
driven primarily by the unconstrained electron density. 
Future density-sensitive diagnostics (e.g., the transauroral $[\mathrm{S\,II}]$ and $[\mathrm{O\,II}]$ lines; \citealt{Holt11, Davies20}) would 
substantially reduce this uncertainty and enable more meaningful comparisons with outflow scaling relations \citep{Fiore17}.

\begin{figure}
\centering
\includegraphics[width=\columnwidth]{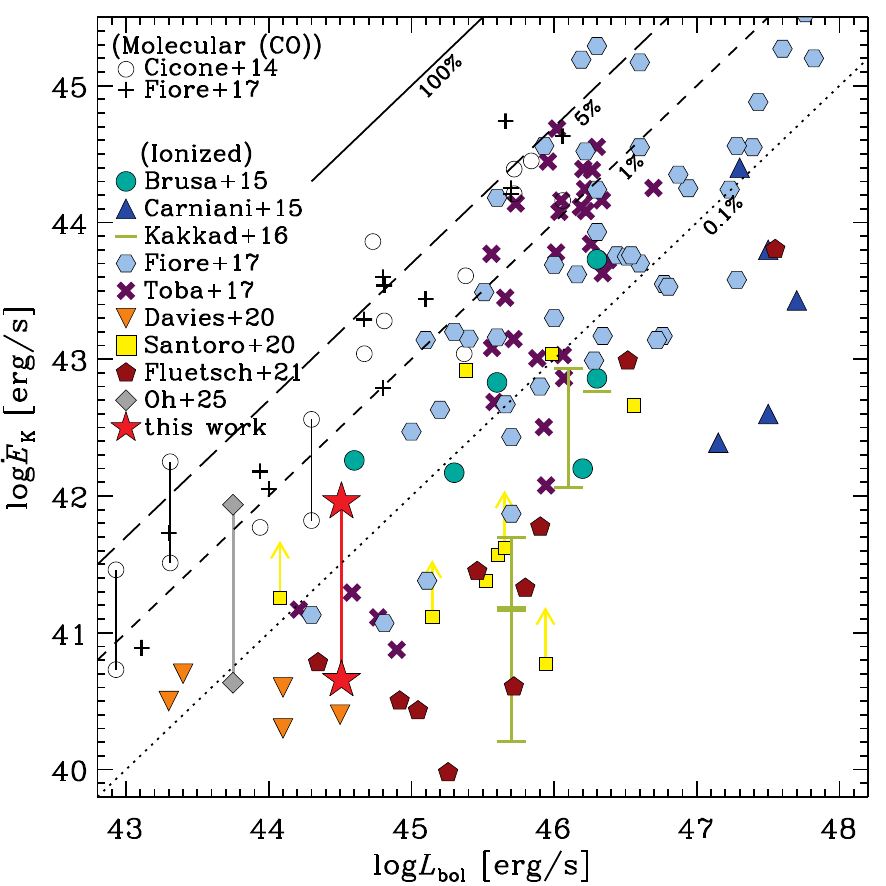}
\caption{Kinetic power of the ionized gas outflow as a function of AGN bolometric luminosity. 
The red stars connected by a vertical line show NGC~6552, 
where the vertical extent reflects the assumed electron density range ($n_{e} = 50$--$1000$~cm$^{-3}$). 
The comparison sample and plotting conventions follow \citet{Oh25}, 
and include molecular (CO) outflows from \citet{Cicone14} and \citet{Fiore17}, 
and ionized outflows from \citet{Brusa15b}, \citet{Carniani15}, \citet{Kakkad16}, \citet{Fiore17}, 
\citet{Toba17}, \citet{Davies20}, \citet{Santoro20}, \citet{Fluetsch21}, and \citet{Oh25}. 
Diagonal lines indicate constant $\dot{E}_{K}/L_{\mathrm{bol}}$ ratios of 100\%, 5\%, 1\%, and 0.1\%. 
NGC~6552 lies near or below the 1\% efficiency line, 
consistent with the expectation that  $[\mbox{O\,{\sc iii}}]$-based measurements trace 
only the ionized phase of a multi-phase outflow.}
\label{fig:Edot_Lbol}
\end{figure}

Figure~\ref{fig:Edot_Lbol} places NGC~6552 on the $\dot{E}_{K}$--$L_{\mathrm{bol}}$ diagram compiled from the literature. 
The vertical extent of the data point reflects the assumed electron density range ($n_{e} = 50$--$1000$~cm$^{-3}$). 
NGC~6552 falls near or below the 1\% coupling efficiency line, 
consistent with other  $[\mbox{O\,{\sc iii}}]$-based ionized outflow measurements at comparable bolometric luminosities.

The most directly comparable measurement is that of \citet{Oh25} for Mrk~766, also shown in Figure~\ref{fig:Edot_Lbol}. 
Although Mrk~766 is Compton-thin, it is a hard X-ray selected AGN of comparable bolometric luminosity ($\log L_{\rm bol} = 43.75$) 
whose outflow energetics were derived with the same KOOLS-IFU instrument, 
\OIII\ tracer, and \citet{CanoDiaz12} formalism adopted here, making it the closest methodologically matched comparison available. 
Its kinetic coupling efficiency ($\dot{E}_{K}/L_{\rm bol} \approx 0.08\%$--$1.53\%$) overlaps with and extends somewhat above that inferred for NGC~6552 ($\approx 0.01\%$--$0.28\%$), 
and both targets fall near or below the 1\% line. 
That a Compton-thick and a Compton-thin source occupy the same low-efficiency region suggests 
this low efficiency is not specific to the Compton-thick regime but instead reflects that \OIII\ traces only a fraction of the multi-phase outflow, 
while direct comparisons with Compton-thick AGN of matched luminosity remain limited by the scarcity of resolved ionized-outflow energetics 
for such heavily obscured sources.

\subsection{Limitations and Caveats}\label{sec:limitations}

Several limitations of the present analysis should be kept in mind.

First, the kinematics--excitation tests rely on a relatively small number of spatial bins that 
satisfy A/N$>3$ in both \OIII\ and \Hb\ ($N_{\rm bins}=15$, Figure~\ref{fig:W80_vs_OIIIHb}). 
A Monte Carlo power analysis using permutation-based Spearman tests 
($2000$ simulated data sets per assumed $\rho$, each tested with $5000$ permutations) 
indicates that, for $N_{\rm bins}=15$ and $\alpha=0.05$, our test achieves 80\% power for $|\rho| \gtrsim 0.7$. 
The null results therefore robustly exclude a strong spatial coupling between either $W_{80}$ or $\Delta v$ and $\log([\mathrm{O\,III}]/\mathrm{H}\beta)$, 
such as would be expected if outflow-driven shocks dominated the line emission at every radius. 
However, moderate correlations ($|\rho| \sim 0.3$--$0.5$) cannot be excluded with the present sample size, 
and deeper observations that improve the S/N of \Hb\ would be valuable for probing subtler trends.

Second, while we have extended the excitation analysis to include $[\mathrm{S\,II}]/\mathrm{H}\alpha$ (Section~\ref{sec:decoupling}), 
a more complete picture would benefit from $[\mathrm{O\,I}]\lambda6300/\mathrm{H}\alpha$ (not reliably detected in our data), 
multi-line mixing sequences, and multi-component fitting to separate narrow and broad kinematic components where warranted.

Third, while our beam-smearing forward model (Section~\ref{sec:W80_origin}) demonstrates that 
a pure rotating disk cannot reproduce the observed $W_{80}$, 
more detailed forward modeling that includes plausible outflow geometries (e.g., biconical models) convolved with the PSF 
would help translate the observed $W_{80}$ into intrinsic outflow velocities and better constrain the outflow structure.

Fourth, tying the \OIII\ core kinematics to the common forbidden-line component, 
which we adopt because the present spectral resolution does not permit an independent shifted component to be fitted in the other forbidden lines, 
could in principle bias the core--shifted decomposition of \OIII. 
Higher-resolution data that allow an untied, multi-line decomposition would permit this assumption to be tested directly.

Finally, the order-of-magnitude outflow energetics presented in Section~\ref{sec:energetics} are dominated by the uncertainty in electron density. 
Measurements of density-sensitive diagnostics beyond the standard $[\mathrm{S\,II}]$ doublet (e.g., the transauroral lines; \citealt{Holt11, Davies20}) 
and robust spatially resolved extinction corrections would substantially tighten these estimates and facilitate more direct comparisons with other gas phases.

Addressing these limitations with deeper and higher spatial-resolution IFU data, together with coordinated 
multi-wavelength analysis, will be crucial for establishing how common kpc-scale, 
multi-phase outflows are in hard X-ray selected obscured AGNs and for quantifying their impact on host galaxies.

\section{Summary and Conclusions}\label{sec:summary}

We have presented KOOLS-IFU optical integral-field spectroscopy of NGC\,6552, 
a hard X-ray selected Compton-thick AGN, obtained on the 3.8\,m Seimei Telescope. 
Using non-parametric \OIII\ kinematic measures and spatially resolved emission-line diagnostics over the central $\sim2$~kpc, 
we investigated whether kinematic broadening is coupled to the local excitation state. 
The \OIII\ non-parametric width $W_{80}$ is broadly elevated across the inner region, 
spanning $\sim$530--830~km~s$^{-1}$, with high-$W_{80}$ values at the smallest projected radii (Figure~\ref{fig:OIII_kinematics}). 
The $W_{80}$ radial profile declines monotonically from the nucleus outward (Figure~\ref{fig:W80_vs_radius}). 
A weighted linear fit provides a good phenomenological description of the trend, 
albeit with substantial scatter ($\chi^{2}_{\nu} \approx 7.4$), suggesting azimuthal variations and/or multiple kinematic components. 
Line-ratio maps of $\log([\mathrm{N\,II}]/\mathrm{H}\alpha)$ and $\log([\mathrm{O\,III}]/\mathrm{H}\beta)$ show 
spatially non-uniform structure within the field of view (Figure~\ref{fig:line_ratio_maps}). 
The spatial footprint of bins that can be robustly classified on the BPT diagram is compact 
and primarily limited by the S/N of H$\beta$, rather than necessarily indicating a sharp physical boundary in excitation. 
All $N_{\rm bins}=15$ bins satisfying A/N$>3$ in all four BPT lines fall in the Seyfert region of 
the \OIIIHb\ versus \NIIHa\ diagram (Figure~\ref{fig:bpt}). 
Across these bins, $W_{80}$ shows no statistically significant correlation with $\log([\mathrm{O\,III}]/\mathrm{H}\beta)$ 
(Spearman $\rho=+0.24$, $p=0.41$; Pearson $r=+0.22$, $p=0.40$; Table~\ref{tab:corr_tests} and Figure~\ref{fig:W80_vs_OIIIHb}). 
This conclusion is unchanged when controlling for projected radius or after subtracting radial trends. 
The velocity asymmetry parameter $\Delta v$ likewise shows no statistically significant correlation with 
$\log([\mathrm{O\,III}]/\mathrm{H}\beta)$ (Spearman $\rho = -0.40$, $p = 0.14$; Pearson $r = -0.39$, $p = 0.15$; Table~\ref{tab:corr_tests}), 
reinforcing the conclusion that kinematic measures are decoupled from excitation diagnostics in this AGN-dominated system. 
Deeper observations that improve the \Hb\ sensitivity and extend the spatial coverage would be valuable for probing whether moderate correlations 
($|\rho| \sim 0.3$--$0.5$) exist below our current detection threshold. 
Together, these results indicate that in NGC~6552 
neither the total \OIII\ line broadening traced by $W_{80}$ nor the line-profile asymmetry traced by $\Delta v$ is 
coupled to the local excitation state.  
Both measures are governed primarily by spatial dynamical structure and 
line-of-sight superposition of multiple kinematic components in this AGN-dominated system. 

We emphasize that this full-sample analysis is the primary result of this work. 
We further note, however, that the full-sample null result may be diluted by bins in which a second kinematic component is not statistically required. 
When the analysis is restricted to the bins for which the two-component fit is most strongly favored by the Bayesian Information Criterion (Appendix~\ref{app:appendix_bic}), 
a positive $W_{80}$--$\log([\mathrm{O\,III}]/\mathrm{H}\beta)$ correlation emerges that survives controlling for projected radius. 
This subsample is small and is defined post hoc, so we regard the coupling as suggestive rather than established, and limited to $W_{80}$, 
since $\Delta v$ remains decoupled in all subsamples. 
It nonetheless indicates that a genuine kinematics--excitation coupling may be present in the bins that most robustly trace the outflow. 
Distinguishing this possibility from a true decoupling will require a larger sample of robustly detected outflow components.

Order-of-magnitude outflow energetics yield a kinetic coupling efficiency of $\dot{E}_{K}/L_{\rm bol} \approx 0.01\%$--$0.28\%$ 
(for $n_{e} = 50$--$1000$~cm$^{-3}$), 
placing NGC~6552 near or below the 1\% efficiency line on the $\dot{E}_{K}$--$L_{\rm bol}$ diagram (Figure~\ref{fig:Edot_Lbol}). 

Combined with recent mid-infrared evidence for outflow signatures in NGC\,6552, 
our KOOLS-IFU results emphasize that a multi-phase, multi-wavelength approach is essential 
for interpreting feedback signatures in hard X-ray selected obscured AGNs. 
Future work will benefit from deeper optical observations that improve H$\beta$ sensitivity, 
expanded excitation diagnostics (e.g., \SIIHa\ and \OIHa), 
and forward modeling that accounts for PSF and outflow geometry, 
enabling more direct constraints on intrinsic outflow velocities and energetics.

\begin{acknowledgments}
We thank the anonymous referee for the careful reading and constructive comments that improved this paper. 
K.O. acknowledges support from the Korea Astronomy and Space Science Institute under the R\&D program 
(Project No.2026-1-831-02), supervised by the Korea AeroSpace Administration, 
and the National Research Foundation of Korea (NRF) grant funded by 
the Korea government (MSIT) (RS-2025-00553982).
Y.U. acknowledges the support from the Kyoto University Foundation.
This work was supported by JSPS KAKENHI Grant Number 24K17104 (S.O.) and JP24KJ1507 (Y.N.).

The Seimei telescope at the Okayama Observatory is jointly operated by Kyoto University and the National Astronomical Observatory of Japan (NAOJ), 
with support from the Optical and Infrared Synergetic Telescopes for Education and Research (OISTER) program.
\end{acknowledgments}





%
\facilities{Seimei:3.8m, Seimei(KOOLS-IFU)}

\software{IRAF \citep{Tody86, Tody93}, 
          Hydra \citep{Barden94, Barden95}, 
          \ppxf\ \citep{Cappellari04},
          \texttt{gandalf} \citep{Sarzi06}
          }


\appendix

\section{Spectral Fitting Examples and Emission-line Quality}\label{app:specfit_examples}

\begin{figure*}
\centering
\includegraphics[height=0.82\textheight, keepaspectratio]{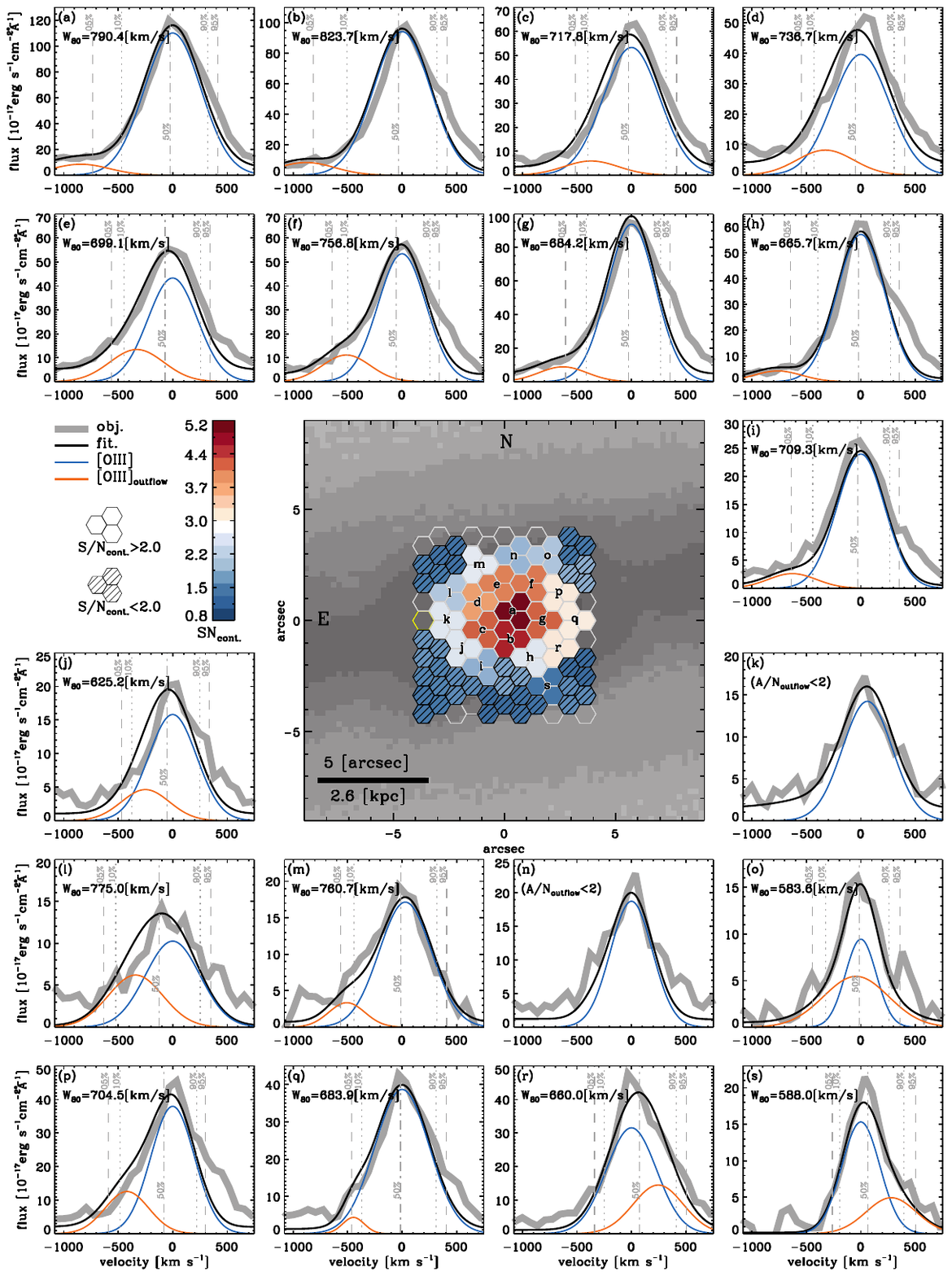}
\caption{
Example \OIII\ line profiles for KOOLS-IFU spatial bins, shown in velocity space. 
Each panel plots the observed spectrum (grey), the full best-fit model including the 
stellar continuum and all emission lines (black), the \OIII\ core component (blue), 
and the \OIII\ shifted component (orange). 
The shifted (outflow) component is shown only for bins with ${\rm A/N}_{\rm outflow}>2$. 
In bins below this threshold a small outflow contribution may still be present in the 
full best-fit model but is not drawn as a separate orange curve. 
The central hexagon map shows the spatial bin labels and the continuum S/N classification used in this appendix. 
Transparent hexagons (outlines only) mark outermost fibers excluded from this work.
The shifted (orange) component is drawn for all bins with 
$\mathrm{A/N_{outflow}} > 2$, which is not identical to the bin-by-bin BIC preference 
shown in Figure~\ref{fig:dBIC_map}. In several bins the single-component model is mildly 
preferred by the BIC despite a formal $\mathrm{A/N_{outflow}} > 2$ detection of the 
shifted component.
}
\label{fig:specfit_in_velocity}
\end{figure*}

Figure~\ref{fig:specfit_in_velocity} presents representative \OIII\ line-profile panels for individual KOOLS-IFU spatial bins, 
displayed in velocity space relative to the systemic \OIII\ wavelength. 
The purpose of this figure is to illustrate how the two-component (\OIII\ core plus shifted component) model describes asymmetric \OIII\ wings, 
and how the adopted quality metrics vary across the field.

We define the local continuum signal-to-noise ratio, ${\rm S/N}_{\rm cont}$, 
as the mean of the observed spectral signal-to-noise measured in two continuum windows near \OIII\ (4890--4920\,\AA\ and 5025--5055\,\AA). 
In the central hexagon map, bins with ${\rm S/N}_{\rm cont}>2$ are color-coded, while bins with ${\rm S/N}_{\rm cont}<2$ are indicated by hatched hexagons.
A small number of fibers at the outermost edge of the array are excluded from our analysis, 
as they both have low continuum S/N and lack adjacent fibers available to form a complete three-fiber bin. 
These fibers carry no measurement and are drawn as transparent (outline-only) hexagons.

The shifted (outflow-related) \OIII\ component is shown only when its Gaussian amplitude-to-noise ratio satisfies ${\rm A/N}_{\rm outflow}>2$. 
In bins that do not meet this criterion, the spectrum is shown without the shifted component, and the panel label indicates ${\rm A/N}_{\rm outflow}<2$.

\begin{figure*}
\centering
\includegraphics[width=\textwidth]{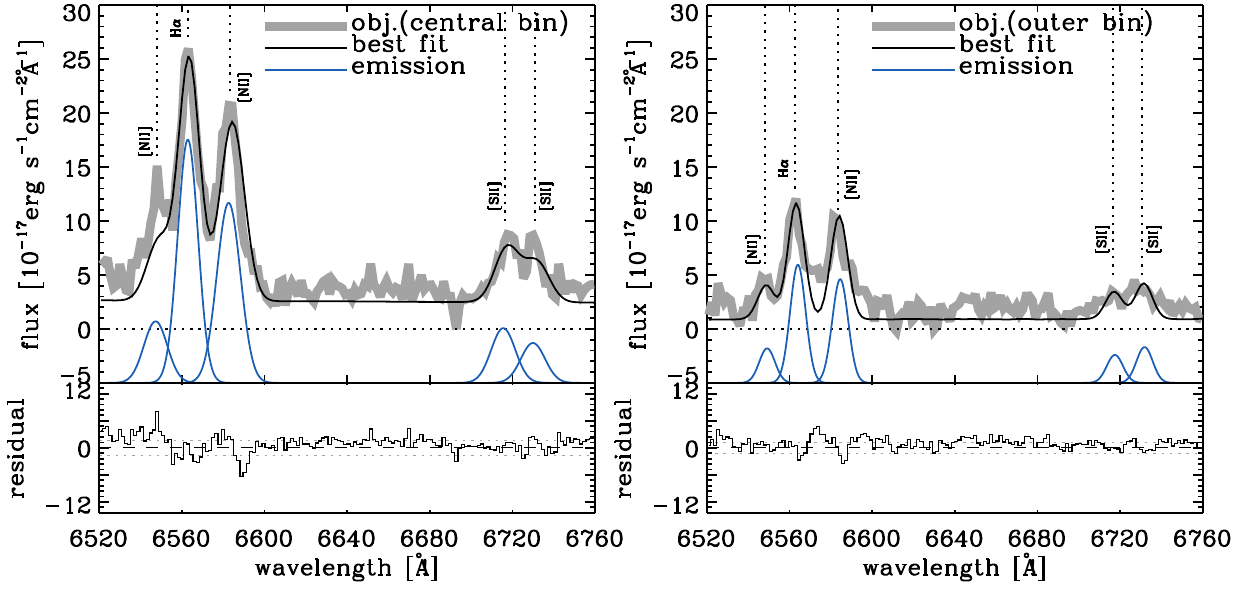}
\caption{
Representative spectral fitting results for the \Ha\ spectral complex regions for the central bin~(a) (left) and an outer bin~(s) (right), 
where the bin labels correspond to those in Figure~\ref{fig:specfit_in_velocity}.
The spectra are shown in the rest-frame wavelength. 
The thick grey shows the observed spectrum, 
the black solid line shows the best-fit model, 
and blue Gaussians show the individual emission-line components. 
The lower panel of each plot shows the fit residual (data minus the full best-fit model), 
with the zero level (dashed) and the $\pm1\sigma$ noise band (dotted) indicated.
These two bins span the range of signal-to-noise across the KOOLS-IFU field and illustrate the quality of the red-spectral-region fitting 
used to construct the \NIIHa\ line-ratio map and to perform the \SIIHa\ 
correlation analysis presented in the main text.
}
\label{fig:specfit_ha}
\end{figure*}

Figure~\ref{fig:specfit_ha} presents representative spectral fitting results for the \Ha\ spectral regions (VPH683 grism) 
for two spatial bins: the central bin~(a) and an outer bin~(s), 
which span the range of signal-to-noise across the KOOLS-IFU field. 
Both panels show the observed spectrum, the best-fit model, and the individual emission-line components. 
The central bin exhibits a prominent \Ha/\NII\ blend that is well described by the simultaneous multi-Gaussian fitting procedure, 
while \SII\ is detected at a lower but usable signal level. 
The outer bin shows reduced overall flux but retains reliable detections of \Ha, \NIIb, 
and \SII, confirming that the shock-sensitive \SIIHa\ diagnostic employed in 
Section~\ref{sec:decoupling} is supported by adequate signal quality across the field.

\section{Selection effects in the $W_{80}$ map}\label{app:w80_selection}

\begin{figure}
\centering
\includegraphics[width=0.65\linewidth]{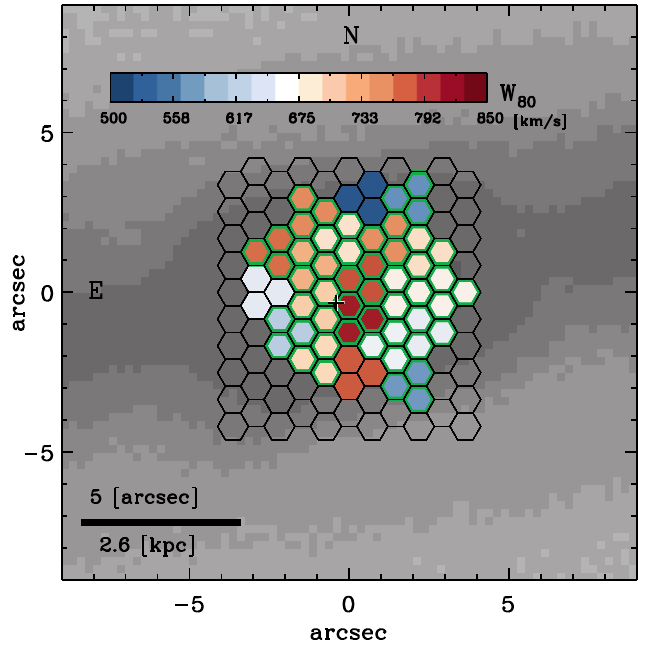}
\caption{
Selection-effect check for the \OIII\ $W_{80}$ map. 
The non-parametric \OIII\ velocity width $W_{80}$ is shown for all spatial bins with reliable \OIII\ detections (A/N$([\mathrm{O\,III}])>3$), 
without requiring detection of the shifted component. 
Unfilled (empty) hexagons indicate bins that do not meet the \OIII\ quality cut. 
Green thick outlines mark bins where the shifted \OIII\ component is detected with ${\rm A/N}_{\rm outflow}>2$ 
(as used for Figure~\ref{fig:OIII_kinematics}). 
}
\label{fig:W80_map_all}
\end{figure}

In the main text (Figure~\ref{fig:OIII_kinematics}), 
we display \OIII\ kinematic maps only for bins where the shifted \OIII\ component is detected with ${\rm A/N}_{\rm outflow}>2$. 
To assess whether this requirement could bias the apparent spatial distribution of large $W_{80}$ values, 
Figure~\ref{fig:W80_map_all} shows a $W_{80}$ map constructed for all bins with reliable \OIII\ measurements (A/N$([\mathrm{O\,III}])>3$), 
without imposing any detection requirement on the shifted component. 
Bins that do not satisfy the \OIII\ quality cut are left unfilled (empty hexagons).

Thick outlines in Figure~\ref{fig:W80_map_all} mark the subset of bins where the shifted component is detected with ${\rm A/N}_{\rm outflow}>2$. 
The overall spatial pattern of enhanced $W_{80}$ remains qualitatively unchanged 
when the shifted-component requirement is removed, 
indicating that the main kinematic features discussed in this work are not driven by the map-selection criterion. 
Rather, the shifted-component selection primarily highlights where the asymmetric wings are detected most robustly.

\section{BIC-Based Robustness Check on the Kinematics--Excitation Correlations}\label{app:appendix_bic}

\begin{figure*}
\centering
\includegraphics[width=0.7\textwidth]{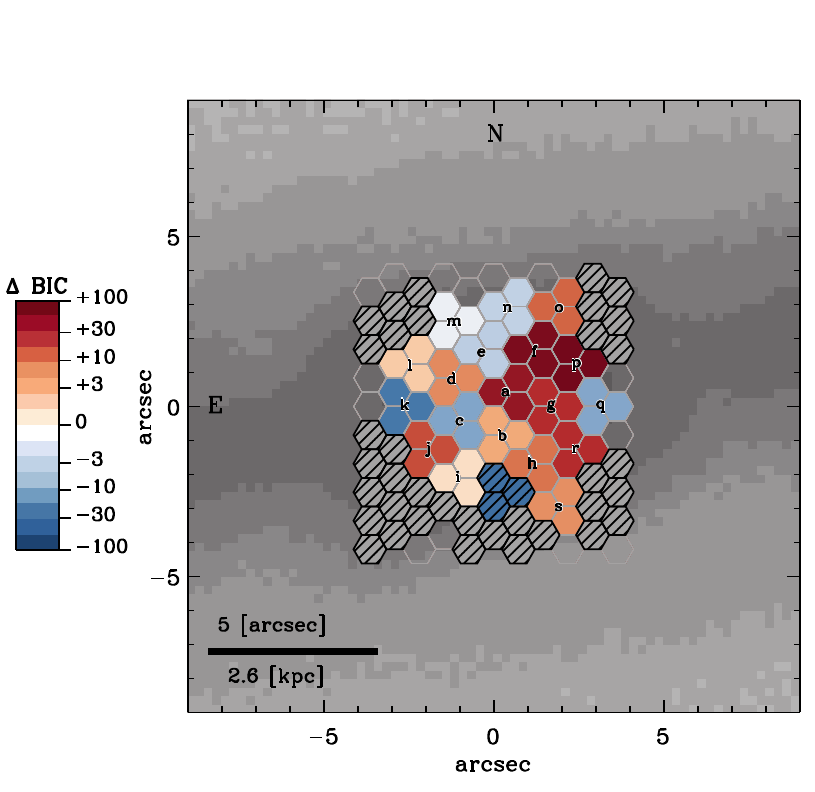}
\caption{$\Delta\mathrm{BIC}$ map of the two-component versus single-component \OIII\ fit across the KOOLS-IFU field. 
Each hexagon shows $\Delta\mathrm{BIC} = \mathrm{BIC}_{1} - \mathrm{BIC}_{2}$ for one spatial bin, 
where positive values favor the two-component (core $+$ shifted) model. 
The color scale uses a signed logarithmic mapping with a cap at $|\Delta\mathrm{BIC}| = 100$ to compress the dynamic range while preserving the sign and rank order. 
Tick marks indicate $\Delta\mathrm{BIC} = \pm 100, \pm 30, \pm 10, \pm 3$, and $0$. 
Bins with $\Delta\mathrm{BIC} > 10$ correspond to ``very strong'' evidence for the two-component model in the \citet{Kass95} scheme and 
are used to define the BIC-strong subsample in Section~\ref{app:appendix_bic_corr}.}
\label{fig:dBIC_map}
\end{figure*}

To assess whether the null kinematics--excitation correlations reported in Section~\ref{sec:results_coupling} 
depend on the choice of outflow-component detection threshold, 
we repeated the correlation tests on subsamples defined by the Bayesian Information Criterion (BIC) 
for the two-component versus single-component \OIII\ fit (Section~\ref{sec:bic}). 
The $\Delta$BIC value of each spatial bin is shown in Figure~\ref{fig:dBIC_map}. 
Following \citet{Kass95}, we define two subsamples in addition to the Original (A/N $>$ 3) sample of Section~\ref{sec:results_coupling}: 
a ``weak'' subsample requiring $\Delta\mathrm{BIC} > 0$, and 
a ``very strong'' subsample requiring $\Delta\mathrm{BIC} > 10$. 
We additionally explore a continuous threshold sweep across $\Delta\mathrm{BIC} \in \{0, 2, 6, 10, 15, 20\}$ 
to characterize the dependence of the correlations on the choice of cut.

\subsection{Correlations on the BIC Subsamples}
\label{app:appendix_bic_corr}

Table~\ref{tab:bic_corr} summarizes the Spearman and partial Spearman correlations on the three \OIIIHb\ subsamples. 
For the Original sample ($N_{\mathrm{bins}} = 15$), neither $W_{80}$ nor $\Delta v$ shows a significant correlation 
with \OIIIHb\ in log-scale, reproducing the result of Table~\ref{tab:corr_tests}. 
In the $\Delta\mathrm{BIC} > 0$ subsample ($N_{\mathrm{bins}} = 11$), 
the $W_{80}$--$\log([\mathrm{O\,III}]/\mathrm{H}\beta)$ correlation strengthens to Spearman $\rho = +0.36$ but remains insignificant ($p = 0.28$), 
and the $\Delta v$ correlation reaches marginal significance ($\rho = -0.55$, $p = 0.083$). 
In the $\Delta\mathrm{BIC} > 10$ subsample ($N_{\mathrm{bins}} = 7$), 
both raw correlations are nominally significant: Spearman $\rho(W_{80}$, $\log([\mathrm{O\,III}]/\mathrm{H}\beta)) = +0.82$ ($p = 0.034$), 
and Spearman $\rho(\Delta v$, $\log([\mathrm{O\,III}]/\mathrm{H}\beta)) = -0.86$ ($p = 0.018$).

\begin{table*}
\centering
\caption{Kinematics--excitation correlations on BIC-defined subsamples of \OIIIHb.}
\label{tab:bic_corr}
\begin{tabular}{lcccc}
\hline\hline
Subsample & $N_{\mathrm{bins}}$ & Test & Statistic & $p$-value \\
\hline
\multicolumn{5}{c}{Original (A/N $>$ 3 in \OIII\ and \Hb)} \\
\hline
$W_{80}$ vs $\log([\mathrm{O\,III}]/\mathrm{H}\beta)$              & 15 & Spearman          & $+0.24$ & 0.41 \\
$W_{80}$ vs $\log([\mathrm{O\,III}]/\mathrm{H}\beta)$ (control $r$) & 15 & Partial Spearman & $-0.17$ & 0.56 \\
$\Delta v$ vs $\log([\mathrm{O\,III}]/\mathrm{H}\beta)$            & 15 & Spearman          & $-0.40$ & 0.14 \\
$\Delta v$ vs $\log([\mathrm{O\,III}]/\mathrm{H}\beta)$ (control $r$) & 15 & Partial Spearman & $-0.035$ & 0.91 \\
\hline
\multicolumn{5}{c}{$\Delta\mathrm{BIC} > 0$ (weak evidence for two components)} \\
\hline
$W_{80}$ vs $\log([\mathrm{O\,III}]/\mathrm{H}\beta)$              & 11 & Spearman          & $+0.36$ & 0.28 \\
$W_{80}$ vs $\log([\mathrm{O\,III}]/\mathrm{H}\beta)$ (control $r$) & 11 & Partial Spearman & $-0.13$ & 0.73 \\
$\Delta v$ vs $\log([\mathrm{O\,III}]/\mathrm{H}\beta)$            & 11 & Spearman          & $-0.55$ & 0.083 \\
$\Delta v$ vs $\log([\mathrm{O\,III}]/\mathrm{H}\beta)$ (control $r$) & 11 & Partial Spearman & $+0.08$ & 0.82 \\
\hline
\multicolumn{5}{c}{$\Delta\mathrm{BIC} > 10$ (very strong evidence for two components)} \\
\hline
$W_{80}$ vs $\log([\mathrm{O\,III}]/\mathrm{H}\beta)$              &  7 & Spearman          & $+0.82$ & 0.034 \\
$W_{80}$ vs $\log([\mathrm{O\,III}]/\mathrm{H}\beta)$ (control $r$) &  7 & Partial Spearman & $+0.84$ & 0.043 \\
$\Delta v$ vs $\log([\mathrm{O\,III}]/\mathrm{H}\beta)$            &  7 & Spearman          & $-0.86$ & 0.018 \\
$\Delta v$ vs $\log([\mathrm{O\,III}]/\mathrm{H}\beta)$ (control $r$) &  7 & Partial Spearman & $-0.40$ & 0.41 \\
\hline
\end{tabular}
\tablecomments{All $p$-values are two-sided permutation-based ($N_{\mathrm{perm}} = 50{,}000$). The Original sample reproduces the corresponding rows of Table~2. The partial Spearman is computed via rank residuals after a linear regression of the rank of each variable on the rank of $r$.}
\end{table*}

\subsection{Robustness Tests}
\label{app:appendix_bic_robust}

We tested the robustness of these BIC-strong correlations using three diagnostics. 
First, a leave-one-out jackknife on the $\Delta\mathrm{BIC} > 10$ subsample yields raw Spearman coefficients 
in the range $+0.71$ to $+0.94$ for $W_{80}$ and $-0.77$ to $-0.94$ for $\Delta v$, 
confirming that no single bin dominates the raw correlations. 
Second, a threshold sweep across $\Delta\mathrm{BIC} \in \{0, 2, 6, 10, 15, 20\}$ shows that 
both correlations strengthen monotonically as the threshold is raised and approach a plateau at the most stringent cuts, 
arguing against a small-sample statistical artifact. 
Third, the partial Spearman correlation controlling for projected radius distinguishes the two kinematic measures: 
$W_{80}$ retains a strong positive partial correlation across all jackknife realizations (range $+0.78$ to $+0.95$), 
whereas $\Delta v$ exhibits a sign-changing partial correlation (range $-0.83$ to $+0.38$), 
indicating that the raw $\Delta v$--$\log([\mathrm{O\,III}]/\mathrm{H}\beta)$ coupling in the BIC-strong subsample is 
largely driven by shared radial dependence rather than a direct local coupling. 
The strong radial decline of $\log([\mathrm{O\,III}]/\mathrm{H}\beta)$ in the BIC-strong subsample (Spearman $\rho = -0.87$) is consistent with this interpretation.

\subsection{Implications for the Main Conclusions}
\label{app:appendix_bic_implications}

The BIC-based robustness check supports the main result of Section~\ref{sec:results_coupling} that 
the velocity asymmetry $\Delta v$ is not coupled to the local excitation state in NGC~6552. 
The $\Delta v$--$\log([\mathrm{O\,III}]/\mathrm{H}\beta)$ decoupling holds across all subsamples once shared radial dependence is controlled for.

A positive $W_{80}$--$\log([\mathrm{O\,III}]/\mathrm{H}\beta)$ coupling, 
by contrast, emerges in the BIC-strong subsample ($\Delta\mathrm{BIC} > 10$, 
$N_{\mathrm{bins}} = 7$), with a partial Spearman coefficient that survives controlling for projected radius (Table~\ref{tab:bic_corr}). 
We do not adopt a fixed bin-number threshold below which such a correlation is disregarded. 
Our caution regarding this subsample reflects three factors. 
It is defined post hoc, its significance was evaluated across a range of $\Delta\mathrm{BIC}$ thresholds 
so that the look-elsewhere effect is non-negligible, and the absolute number of bins is small. 
The emergence of the correlation is nonetheless informative. 
Several bins that enter the full sample with a formal A/N$>2$ detection of the shifted component are preferred 
as single-component by the BIC, so the full-sample correlation is averaged over bins in which a second kinematic component is not statistically required. 
The full-sample null result of Section~\ref{sec:results_coupling} is therefore the conservative outcome, 
while the BIC-strong subsample isolates the bins that most robustly trace the outflow.

The primary kinematics--excitation test of this work is the full-sample test of Section~\ref{sec:results_coupling}, which yields a null result. 
The positive $W_{80}$--$\log([\mathrm{O\,III}]/\mathrm{H}\beta)$ coupling discussed here is a secondary result that emerges only in the small BIC-strong subsample. 
If it is genuine, it would be consistent with a picture in which the bins that most robustly trace the outflow also sample the most actively AGN-photoionized gas. 
Confirming whether this coupling is real rather than a small-sample fluctuation requires a larger number of bins with robustly detected outflow components, 
which deeper observations would provide.

\section{Spatial Distribution of the Balmer Decrement}
\label{sec:appendix_balmer}

\begin{figure*}
\centering
\includegraphics[width=0.7\textwidth]{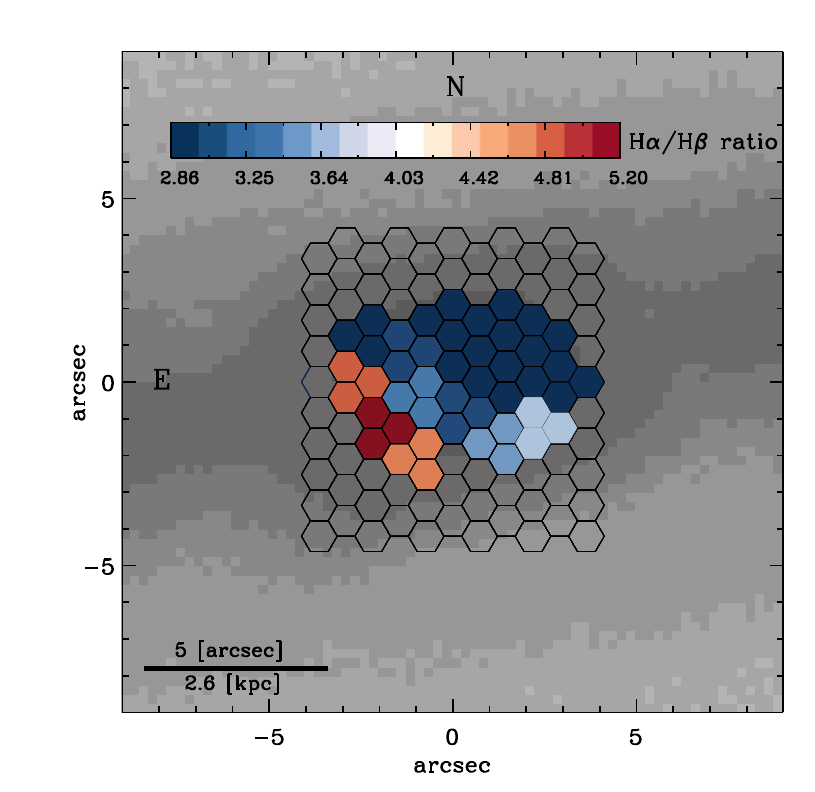}
\caption{Spatially resolved Balmer decrement \Ha/\Hb\ map of NGC~6552 overlaid on the Pan-STARRS1 $r$-band image. 
Color-coded hexagonal bins ($N=15$) satisfy A/N $> 3$ in \OIII, \Ha, and \Hb, 
consistent with the main kinematics--excitation sample of Section~\ref{sec:results_coupling}. 
The \Ha/\Hb\ ratio ranges from $\sim 2.86$ (the Case~B recombination value at $T = 10^4$~K; \citealt{Osterbrock06}, indicating no internal extinction) 
to $\sim 5.14$, corresponding to a maximum internal reddening of $E(B-V)_{\rm int} \approx 0.50$~mag using the \citet{Calzetti00} attenuation curve. 
Seven of the 15 bins (47 percent) have \Ha/\Hb\ $\le 2.86$ and are assigned $E(B-V)_{\rm int} = 0$ (consistent with measurement noise). 
Unfilled hexagons mark bins that fail the A/N $> 3$ quality cut. 
The bar at lower-left indicates the angular and physical scales. 
The N and E directions are labeled.}
\label{fig:balmer_decrement_map}
\end{figure*}

To complement the quantitative analysis of differential extinction presented in Section~\ref{sec:decoupling}, 
Figure~\ref{fig:balmer_decrement_map} shows the spatially resolved Balmer decrement \Ha/\Hb\ map 
for the $N=15$ bins satisfying A/N $> 3$ in \OIII, \Ha, and \Hb, 
the same sample used in the main kinematics--excitation correlation tests in Section~\ref{sec:results_coupling}.

The map shows that the majority of bins, including those near the nucleus, 
exhibit \Ha/\Hb\ values consistent with the Case~B recombination limit ($\sim 2.86$), 
indicating no measurable internal extinction. 
Seven of the 15 bins (47 percent) have \Ha/\Hb\ $\le 2.86$. 
Higher \Ha/\Hb\ values are concentrated in a localized region to the southeast of the nucleus at $r \sim 1.2$~kpc, 
where \Ha/\Hb\ reaches $\sim 5.14$ (corresponding to $E(B-V)_{\rm int} \approx 0.50$~mag using the \citet{Calzetti00} attenuation curve).

This spatial pattern is consistent with the picture, discussed in Section~\ref{sec:decoupling}, 
in which the optical line of sight preferentially samples less-obscured narrow-line region gas, 
with the most highly ionized inner gas being attenuated by the Compton-thick obscuration intrinsic to NGC~6552 
($\log(N_{\rm H}/{\rm cm}^{-2}) = 24.05$; \citealt{Ricci17}). 
The marginal negative correlation between \Ha/\Hb\ and $\log([\mathrm{O\,III}]/\mathrm{H}\beta)$ 
(Spearman $\rho = -0.43$, $p = 0.11$) reflects this shared dependence on projected radius 
rather than a direct differential reddening effect, 
as confirmed by the small dynamic range in internal extinction (median $E(B-V)_{\rm int} = 0.04$~mag) 
and the opposite sign of the correlation compared to what differential extinction alone would produce.

\bibliography{reference}{}
\bibliographystyle{aasjournalv7}



\end{document}